\documentclass[nofootinbib,secnumarabic,amssymb,superscriptaddress, aps, pre]{revtex4}

\usepackage{amsmath}
\usepackage{mathtools}
\usepackage{graphicx}
\usepackage[dvipsnames]{xcolor}
\usepackage{epsfig}
\usepackage{mathrsfs}
\usepackage{amsbsy}
\usepackage{verbatim}

\usepackage[colorlinks,citecolor=blue]{hyperref}
\usepackage[capitalise,nameinlink]{cleveref}
\hypersetup{colorlinks=true, linkcolor=blue}
\hypersetup{urlcolor=blue}
\usepackage{caption,subcaption}
\usepackage{cases}
\usepackage{empheq}
\usepackage{orcidlink}

\newcommand{\ba}{\begin{align}}
\newcommand{\ea}{\end{align}}
\newcommand{\nn}{\nonumber}
\newcommand{\be}{\begin{equation}}
\newcommand{\ee}{\end{equation}} 
\newcommand{\bea}{\begin{eqnarray}}
\newcommand{\eea}{\end{eqnarray}}
\def\nn{\nonumber}

\usepackage{latexsym,oldgerm}
\def\be{\begin{equation}}
\def\ee{\end{equation}}
\def\bea{\begin{eqnarray}}
\def\eea{\end{eqnarray}}
\def\nn{\nonumber}

\allowdisplaybreaks

\begin{document}
\title{Reciprocal theorem for linear poro-viscoelastic materials }

\author{Moslem Moradi}
\affiliation{Department of Applied Physical Sciences, University of North Carolina at Chapel Hill, Chapel Hill, NC 27599-3250}
\affiliation{Department of Chemical Engineering, University of Pittsburgh, Pittsburgh, PA 15261}
\author{Wenzheng Shi}
\affiliation{Department of Applied Physical Sciences, University of North Carolina at Chapel Hill, Chapel Hill, NC 27599-3250}
\author{Ehssan Nazockdast}
\affiliation{Department of Applied Physical Sciences, University of North Carolina at Chapel Hill, Chapel Hill, NC 27599-3250}

\email[Email: ]{ehssan@email.unc.edu}

\date{\today}
\begin{abstract}
\vspace*{0.55cm}
\subsection*{Abstract}
\noindent 
In studying the transport of particles and inclusions in multi-phase systems 
we are often interested in integrated quantities such as the total force and the net velocity of the particles. 
The divergence form of momentum equation in linear viscous and elastic materials allows for rewriting these equations as integrals over the involved interfaces, providing a fast method of calculating the force and velocity at the interface without the need to compute the detailed velocity and displacements fields. Central to these integral representation is the reciprocal theorem, where a known solution to the boundary value problem is used to compute the integrated quantities in the same geometry but subjected to different boundary conditions.
Here, we derive a reciprocal formulation for linear poro-viscoelastic (PVE) materials, which are composed of a linear compressible viscoelastic phase, i.e. the network phase, permeated by a viscous fluid.
As an application of the reciprocal theorem, we analytically calculate the time-dependent net force on a rigid stationary sphere in response to point-forces applied to the elastic network and Newtonian fluid phases of a PVE material. We show that the net force on the sphere in response to a point-force in the fluid phase evolves over timescales that are independent of the distance of the point-force to the sphere; in comparison, when the point-force is applied to the network phase the timescale for force development becomes distance-dependent. We discuss how in both cases these relaxation times are related to the physical timescales that are determined by mechanical properties of both phases --such as the network's Poisson ratio, permeability and shear modules, and the fluid viscosity-- as well as geometric factors, including the size of the spherical inclusion and its distance from point-forces. The reciprocal theorem presented here can be applied to a wide range of problems involving the transport of cells,  organelles and condensates in biological systems composed of filamentous networks permeated by viscous fluids. 
 
\end{abstract}
\maketitle

\section{Introduction}
Multiphase systems composed of deformable solid structures permeated by fluids are found in many natural and industrial settings, including flow and mass transport in soils and rocks \citep{cheng2016poroelasticity}, and polymer gels \citep{doi2009gel}. 
Many biological materials are composed of  flexible filamentous networks that are permeated by a fluid-like phase \citep{alberts2003molecular}, including tissues \citep{cowin2007tissue}, the cell cytoskeleton \citep{howard2001mechanics} and the nucleus \citep{zidovska2020rich, hobson2021survey}. This is the primary application of interest for the current work. 
The function of many biological systems strongly depends on their mechanical properties \citep{ethier2007introductory}, and many pathologies coincide with changes in mechanical properties. Hence, cell mechanics is increasingly used as a biomarker for studying cellular processes \citep{fletcher2010cell} as well as detecting and separating diseased cells \citep{Darling2015}. 

In most of the mentioned applications, the transport occurs over lengthscales significantly larger than the building blocks of the solid phase, which makes continuum modeling computationally more feasible than particle simulations. Different homogenization techniques, including volume averaging, mixture theory and asymptotic  expansions \cite{whitaker1998method, battiato2019theory}, have been used to describe the coupled mechanics of solid and fluid phases in continuum scale.  Perhaps the most well-known example is Biot  theory of poroelasticity \citep{biot1941general, biot1955theory}, where the fluid and elastic phases are coupled through a friction term that is linearly proportional to the relative velocity of the phases. In Biot theory, the fluid shear stresses are assumed to be negligible  and the fluid momentum equation is described by Darcy's equation. For the reasons outlined in the next two paragraphs, shear stresses are included in the current work and Brinkman equation is used instead for modeling fluid momentum equation. 

The semi-flexible filament networks in biological materials only constitute a very small volume fraction of the system. Scaling analysis shows that Brinkman equation is a more accurate description of the fluid flows than Darcy's equation, when the volume fraction of the network (solid phase hereafter referred to as network) is very small \citep{auriault2009domain, levy1983fluid}.  
Furthermore, in many instances, e.g. in the cell nucleus \citep{zidovska2020rich}, the underlying poroelastic materials can be highly heterogeneous, made of pure fluid domains with nearly infinite permeability, to distinct domains that are packed with the network phase. Brinkman equation has the advantage of asymptoting to correct limits of Stokes flow in network-free phases and Darcy's equation in small permeability. It also provides a natural framework for imposing the physically consistent boundary conditions at interfaces, where fluid shear forces cannot be ignored \citep{carrillo2019darcy, carrillo2020multiphase}. 

Another example, where Brinkman equation provides clear advantages over Darcy's equation, is in studying the time-dependent response of the biological materials in microrheological techniques. 
As discussed in details in \citep{levine2000one, levine2001response} and \cite{moradi2022general}, at times smaller than $\tau^\star=a^2\xi/(2G+\lambda)$, the network and fluid phases co-move and the viscoelastic response of the PE material is determined only by the network's shear modulus and the fluid viscosity. Here, $G$ and $\lambda$ are the first and second Lamé coefficients of the network, and $\xi$ is the friction coefficient that relates the frictional body force to the relative motion of the fluid and the network phase in each phase. In this limit Generalized Stokes-Einstein (GSE) relationship can be used to determine the rheology. These shear relaxation modes are absent in Biot's theory. In comparison, fluid flows in response to network compressibility develop at $t>\tau^\star$ and lead to a slower distance-dependent relaxation dynamics that is captured using Biot's theory, and not in GSE framework \citep{moradi2022general}. Experimental and 
modeling studies suggest that these osmotic effects may be a determining factor in cellular processes such as blebbing \citep{charras2005non}. 
Using Brinkman equation allows us to study the dynamics across different timescales and lengthscales.  

The main objective in continuum simulations is to compute the distribution of stress and velocity fields in the fluid and network phases in a control volume that is bounded by (simply or multiply-connected) interfaces. 
We assume that the solid and fluid phases can be described as linear viscoelastic materials,  which leads to two coupled linear homogeneous elliptic PDEs and a mass conservation. Finite element \citep{simon1992multiphase} and immersed boundary  \citep{strychalski2015poroelastic} methods have been used in the past to solve these PDEs. 
A similar mathematical problem arises in Stokes flow and linear elasticity, when studying the transport of particles/droplets. 
In both cases, the linear homogeneous elliptic form of the momentum equation for velocity (Stokes) and displacement (elasticity), allows one to develop
robust mathematical formulations that can be used to obtain analytical solutions
to a variety of problems involving simple geometries and to develop efficient numerical methods such as boundary integral and singularity methods \citep{kim2013microhydrodynamics, pozrikidis2002practical} for complex geometries.
Our main objective is to develop a similar set of tools for linear poro-viscoelastic (PVE) materials with non-negligible viscous shear forces, with applications to intracellular and extracellular assemblies of filaments and biopolyemrs. 
In an earlier study \citep{moradi2022general}, we presented the general solutions for PVE materials.
Here, we develop another piece of this mathematical framework by formulating the reciprocal theorem for this class of materials.  

The reciprocal theorem appears in different areas of physics and engineering, including optics, acoustics, elasticity, fluids and electromagnetism, which are
surveyed in \citep{masoud2019reciprocal}. 
The main idea in reciprocal theorem is that the solutions to an auxiliary problem under a given set of boundary conditions (BCs) can be used to obtain the solutions under another more complex set of BCs  without having to solve the boundary value problem. The auxiliary problem must be sufficiently simple to obtain closed-form solutions.
Because of the linearity of the equations, the reciprocal theorem implies that the relation between kinetics and kinematics is linear and the response functions must be symmetric \citep{lauga2009hydrodynamics}. 
Apart from its mathematical elegance, the reciprocal theorem is useful in determining the motion of swimmers in Newtonian and viscoelastic fluids \citep{li2021microswimming, elfring2015note,lauga2009hydrodynamics}, and finding analytical solutions in simple geometries \citep{selvadurai2000application}. 
The integral representation for the unknown variables are also the restatement of the governing equations from a three-dimensional PDE to a two-dimensional integral equation for unknown densities over the boundary of the fluid domain, which forms the basis of boundary integral numerical methods \citep{pozrikidis2002practical}.

The paper is organized as follows.
In section \S \ref{sec:reciprocal}, we present the governing equations for linear PVE materials followed by the derivation for the reciprocal theorem. In section \S \ref{sec:results} we use the reciprocal theorem to derive analytical expressions for the net force on a rigid stationary sphere in response in point-forces in the Newtonian fluid phase and the linear compressible elastic network phase of a linear PVE material. 
The summary of our results and brief discussions on other applications of the reciprocal theorem are discussed in section \S \ref{sec:summary}.

\section{Governing equations and the reciprocal theorem}
\label{sec:reciprocal}
The conservation equations for a two-phase system composed of a dilute network phase permeated by a fluid phase are: 
\begin{subequations} \label{eq:PE}
\begin{align}
 \mathrm{Mass \,conservation:} & \quad \nabla \cdot \left(\phi \boldsymbol{v}_n+(1-\phi)\boldsymbol{v}_f\right)=0 , \label{eq:PEa} \\
 \mathrm{Momentum\, Eq.\, for\, the\, fluid:} & \quad \nabla \cdot \left(\boldsymbol{\sigma}_f-(1-\phi) p \mathbf{I}\right)+\xi \left(\boldsymbol{v}_f-\boldsymbol{v}_n\right) +\mathbf{f}_f=\mathbf{0} ,  \label{eq:PEb} \\
  \mathrm{Momentum\, Eq.\, for\, the\, network:} & \quad \nabla \cdot \left(\boldsymbol{\sigma}_n-\phi p \mathbf{I}\right)-\xi \left(\boldsymbol{v}_f-\boldsymbol{v}_n\right)+\mathbf{f}_n=\mathbf{0},  \label{eq:PEc}
\end{align}
\end{subequations}
where subscripts $n$ and $f$ denote network and fluid phases, $\phi \ll 1$ is the volume fraction of the network phase, $\boldsymbol{\sigma}_n$ is the network stress, $p$ is the pressure 
that ensures fluid incompressibility constraint; $\xi$ is the friction constant per unit volume of the material, $\mathbf{f}_f$, and $\mathbf{f}_n$ are the external body forces on the network and fluid phases, respectively, and $\mathbf{I}$ is the identity matrix. 
We take the volume fraction of the network phase, $\phi$, to remain constant in space and time. Thus, the compressibility of the network leads to nonzero divergence of the fluid velocity field: $\nabla \cdot \boldsymbol{v}_f =(\phi/(1-\phi))\nabla \cdot \boldsymbol{v}_n \neq 0$. However, this apparent fluid compressibility is not associated with  changes in fluid density and arises from fluid sources and sinks in response to changes in volume of the network phase. As such, the Newtonian fluid stress does not include the isotropic term that scales with $\eta_b \left(\nabla \cdot \boldsymbol{v}_f\right)$, where $\eta_b$ is the bulk viscosity. We use a general linear (VE) isotropic constitutive equation (CE) to describe the traceless component of the fluid stress: 
\begin{equation}
{\boldsymbol{\sigma}}_f =  \int_{0}^{t}  G_f(t-t^\prime)  \left( \nabla {\boldsymbol{v}}_f (t^\prime)+ \nabla {\boldsymbol{v}}_f^{T} (t^\prime) \right) \mathrm{d} t^\prime, 
\label{eq:FCE}
\end{equation} 
where $G_f(t)$ is the fluid's shear modulus. 
Similarly, we model the network stress, $\boldsymbol{\sigma}_n$, using a general linear VE constitutive equation: 
\begin{eqnarray}
&&  {\boldsymbol{\sigma}}_n =  \int_{0}^{t} \mathbf{C}(t-t^\prime) : \left( \nabla {\boldsymbol{v}}_n (t^\prime)+ \nabla {\boldsymbol{v}}_n^{T} (t^\prime) \right)dt^\prime, 
\label{eq:PECE}
\end{eqnarray} 
where $\mathbf{C}(t)$ is the symmetric fourth-rank stiffness tensor, and '':'' is double dot product. 
Taking Laplace transform of Eqs. \eqref{eq:FCE} and \eqref{eq:PECE} gives
$\tilde{\boldsymbol{\sigma}}_f=\tilde{\eta}(s) \left(\nabla \tilde{\boldsymbol{v}}_f+\nabla \tilde{\boldsymbol{v}}^T_f\right)$ and 
$\tilde{\boldsymbol{\sigma}}_n=\tilde{\mathbf{C}}(s) :\left(\nabla \tilde{\boldsymbol{v}}_n+\nabla \tilde{\boldsymbol{v}}^T_n\right)$, where superscripts $\sim$ denote variables in Laplace space, and $\tilde{\eta}(s)=\mathcal{L}(G_f(t))$, where $\mathcal{L}$ is the Laplace transform operator. 
Taking Laplace transform of Eqs. \eqref{eq:PE} and using convolution theorem we get:
\begin{subequations}
\begin{align}
  &  \nabla \cdot \left( {\tilde{\boldsymbol{v}}}_n \phi +  {\tilde{\boldsymbol{v}}}_f (1- \phi)   \right) =0,\\
  &  \nabla \cdot \tilde{\boldsymbol{\Sigma}}_f+ \xi (\tilde{\boldsymbol{v}}_n - \tilde{\boldsymbol{v}}_f  ) =-\tilde{\mathbf{f}}_f, 
    \label{eq:PEfLaplace1b}\\
&    \nabla \cdot   \tilde{\boldsymbol{\Sigma}}_n - \xi (\tilde{\boldsymbol{v}}_n - \tilde{\boldsymbol{v}}_f  ) =-\tilde{\mathbf{f}}_n \label{eq:PEnLaplace1c}, 
\end{align}
\end{subequations} 
where $\tilde{\boldsymbol{\Sigma}}_f= \tilde{\boldsymbol{\sigma}}_f-(1-\phi)\tilde{p}\mathbf{I}$ and 
$\tilde{\boldsymbol{\Sigma}}_n= \tilde{\boldsymbol{\sigma}}_n-\phi\tilde{p}\mathbf{I}$ are 
the total stresses in the fluid and network phases in Laplace space. 
We take the superscripts $(1)$ and $(2)$ to denote two independent solutions to the above system of equations, i.e. $( {\tilde{\boldsymbol{v}}}_f^{(1)} , \tilde{\boldsymbol{\Sigma}}_f^{(1)}  , \tilde{\mathbf{f}}_f^{(1)} )$, and $( {\tilde{\boldsymbol{v}}}_f^{(2)}  , \tilde{\boldsymbol{\Sigma}}_f^{(2)}  , \tilde{\mathbf{f}}_f^{(2)} )$; then, we take the dot product of Eqs \eqref{eq:PEfLaplace1b} and \eqref{eq:PEnLaplace1c}, with  $ \tilde{\boldsymbol{v}}_f^{(2)}$ and $ \tilde{\boldsymbol{v}}_n^{(2)}$, respectively, to get: 
\begin{subequations}
\begin{align}
     \tilde{\boldsymbol{v}}_f^{(2)}\cdot \left[\nabla \cdot  \tilde{\boldsymbol{\Sigma}}^{(1)}_f+\xi \left( \tilde{\boldsymbol{v}}^{(1)}_n-  \tilde{\boldsymbol{v}}^{(1)}_f \right) \right]&= -  \tilde{\boldsymbol{v}}_f^{(2)}\cdot  \tilde{\mathbf{f}}_f^{(1)},      \label{eq:Recip1a}\\
     \tilde{\boldsymbol{v}}_n^{(2)}\cdot \left[\nabla \cdot  \tilde{\boldsymbol{\Sigma}}_n^{(1)}-\xi \left(  \tilde{\boldsymbol{v}}_n^{(1)}-  \tilde{\boldsymbol{v}}_f^{(1)}\right) \right]&= - \tilde{\boldsymbol{v}}_n^{(2)}\cdot  \tilde{\mathbf{f}}_n^{(1)}.  \label{eq:Recip1b}
\end{align}
\end{subequations}
We repeat the same process, using solution $( {\tilde{\boldsymbol{v}}}_f^{(2)}  , \tilde{\boldsymbol{\Sigma}}_f^{(2)}  , \tilde{\mathbf{f}}_f^{(2)} )$ of Eqs. \eqref{eq:PEfLaplace1b}, and \eqref{eq:PEnLaplace1c} and multiplying  them 
by $\boldsymbol{v}_f^{(1)}$ and $\boldsymbol{v}_n^{(1)}$, respectively, to get:
\begin{subequations}
\begin{align}
     \tilde{\boldsymbol{v}}_f^{(1)}\cdot \left[\nabla \cdot  \tilde{\boldsymbol{\Sigma}}^{(2)}_f+\xi \left( \tilde{\boldsymbol{v}}^{(2)}_n-  \tilde{\boldsymbol{v}}^{(2)}_f \right) \right]&= - \tilde{\boldsymbol{v}}_f^{(1)}\cdot  \tilde{\mathbf{f}}_f^{(2)},    \label{eq:Recip2a}\\
     \tilde{\boldsymbol{v}}_n^{(1)}\cdot \left[\nabla \cdot  \tilde{\boldsymbol{\Sigma}}_n^{(2)}-\xi \left(  \tilde{\boldsymbol{v}}_n^{(2)}-  \tilde{\boldsymbol{v}}_f^{(2)}\right) \right]&= - \tilde{\boldsymbol{v}}_n^{(1)}\cdot  \tilde{\mathbf{f}}_n^{(2)}.   \label{eq:Recip2b}
\end{align}
\end{subequations}
Subtracting Eq. \eqref{eq:Recip1a} with \eqref{eq:Recip2a}, and also Eq. \eqref{eq:Recip1b} with \eqref{eq:Recip2b} gives:
\begin{subequations} \label{Eq12}
\begin{align}
 & \nabla \cdot  \left(  \tilde{\boldsymbol{\Sigma}}_f^{(1)} \cdot  \tilde{\boldsymbol{v}}_f^{(2)}-  \tilde{\boldsymbol{\Sigma}}_f^{(2)} \cdot  \tilde{\boldsymbol{v}}_f^{(1)}\right)+\xi\left[  \tilde{\boldsymbol{v}}_f^{(2)}\cdot (  \tilde{\boldsymbol{v}}^{(1)}_n-  \tilde{\boldsymbol{v}}^{(1)}_f)-  \tilde{\boldsymbol{v}}_f^{(1)}\cdot ( \tilde{\boldsymbol{v}}^{(2)}_n-  \tilde{\boldsymbol{v}}^{(2)}_f)\right] +(1-\phi)\left(  \tilde{p}^{(1)}\nabla \cdot  \tilde{\boldsymbol{v}}_f^{(2)}-  \tilde{p}^{(2)}\nabla \cdot  \tilde{\boldsymbol{v}}_f^{(1)}\right)  \nn \\  
& \qquad    =  \tilde{\boldsymbol{v}}_f^{(1)}\cdot  \tilde{\mathbf{f}}^{(2)}_f-  \tilde{\boldsymbol{v}}_f^{(2)}\cdot  \tilde{\mathbf{f}}^{(1)}_f  \label{Eq12a}\\
& \nabla \cdot \left(  \tilde{\boldsymbol{\Sigma}}_n^{(1)} \cdot  \tilde{\boldsymbol{v}}_n^{(2)}- \tilde{\boldsymbol{\Sigma}}_n^{(2)} \cdot  \tilde{\boldsymbol{v}}_n^{(1)}\right)-\xi\left[  \tilde{\boldsymbol{v}}_n^{(2)}\cdot (  \tilde{\boldsymbol{v}}^{(1)}_n-  \tilde{\boldsymbol{v}}^{(1)}_f)-  \tilde{\boldsymbol{v}}_n^{(1)}\cdot (  \tilde{\boldsymbol{v}}^{(2)}_n-  \tilde{\boldsymbol{v}}^{(2)}_f)\right]  +\phi\left(  \tilde{p}^{(1)}\nabla \cdot  \tilde{\boldsymbol{v}}_n^{(2)}-  \tilde{p}^{(2)}\nabla \cdot  \tilde{\boldsymbol{v}}_n^{(1)}\right)   \nn \\  
&  \qquad    =  \tilde{\boldsymbol{v}}_n^{(1)}\cdot  \tilde{\mathbf{f}}^{(2)}_n-  \tilde{\boldsymbol{v}}_n^{(2)}\cdot  \tilde{\mathbf{f}}^{(1)}_n . \label{Eq12b}
\end{align}
\end{subequations}
Here, we have used the identities $  \tilde{\boldsymbol{v}} \cdot (\nabla \cdot \tilde{\boldsymbol{\Sigma}})=\nabla \cdot \left( \tilde{\boldsymbol{\Sigma}}\cdot \tilde{\boldsymbol{v}} \right)- \tilde{\boldsymbol{\Sigma}} :\nabla \tilde{\boldsymbol{v}} $. 
and $ \tilde{\boldsymbol{e}}^{(1)}:\nabla \tilde{\boldsymbol{v}}^{(2)}= \tilde{\boldsymbol{e}}^{(2)}:\nabla \tilde{\boldsymbol{v}}^{(1)}=\frac{1}{2}  \tilde{\boldsymbol{e}}^{(1)}:
 \tilde{\boldsymbol{e}}^{(2)}$,
where $\tilde{\boldsymbol{e}}= \frac{1}{2}\left(\nabla \tilde{\boldsymbol{v}}+\nabla \tilde{\boldsymbol{v}}^T \right)$ is the rate of strain tensor.
Adding Eqs. \eqref{Eq12a} and \eqref{Eq12b}, directly cancels the second terms on the left hand side, and using continuity equation (($1-\phi)\nabla \cdot \tilde{\boldsymbol{v}}_f+\phi\nabla \cdot \tilde{\boldsymbol{v}}_n =0 $) cancels the terms involving $\tilde{p}^{(1,2)}_{f,n}\left(\nabla \cdot \tilde{\boldsymbol{v}}_{f,n}^{(1,2)}\right)$. The final result is the differential form of the reciprocal relations:
\begin{equation}
    \nabla \cdot \left( \tilde{\boldsymbol{\Sigma}}_f^{(1)} \cdot \tilde{\boldsymbol{v}}_f^{(2)}+ \tilde{\boldsymbol{\Sigma}}_n^{(1)} \cdot \tilde{\boldsymbol{v}}_n^{(2)}\right)+\tilde{\boldsymbol{v}}_f^{(1)}\cdot \tilde{\mathbf{f}}^{(2)}_f+
 \tilde{\boldsymbol{v}}_n^{(1)}\cdot \tilde{\mathbf{f}}^{(2)}_n=  \nabla \cdot \left( \tilde{\boldsymbol{\Sigma}}_f^{(2)} \cdot \tilde{\boldsymbol{v}}_f^{(1)}+ \tilde{\boldsymbol{\Sigma}}_n^{(2)} \cdot \tilde{\boldsymbol{v}}_n^{(1)}\right)+
 \tilde{\boldsymbol{v}}_f^{(2)}\cdot \tilde{\mathbf{f}}^{(1)}_f+ \tilde{\boldsymbol{v}}_n^{(2)}\cdot \tilde{\mathbf{f}}^{(1)}_n.           \label{eq:RT}
\end{equation}

Next, we select a control volume $V$, that is bounded by a closed (simply- or multiply-connected) surface $S=S_p+S_\infty$; see figure \ref{fig1}. 
After integrating Eq.~\eqref{eq:RT} over $V$ and using the divergence theorem we get
\begin{eqnarray}
&& \int_{S_p} \left( \tilde{\boldsymbol{\Sigma}}_f^{(1)} \cdot \tilde{\boldsymbol{v}}_f^{(2)} + \tilde{\boldsymbol{\Sigma}}_n^{(1)} \cdot \tilde{\boldsymbol{v}}_n^{(2)}\right)\cdot  \mathbf{n} \, \mathrm{d}S+\int_V \left( \tilde{\boldsymbol{v}}_f^{(1)}\cdot \tilde{\mathbf{f}}^{(2)}_f+
 \tilde{\boldsymbol{v}}_n^{(1)}\cdot \tilde{\mathbf{f}}^{(2)}_n\right) \mathrm{d}V=   \int_{S_p} \left( \tilde{\boldsymbol{\Sigma}}_f^{(2)} \cdot \tilde{\boldsymbol{v}}_f^{(1)}+ \tilde{\mathbf{\Sigma}}_n^{(2)} \cdot \tilde{\boldsymbol{v}}_n^{(1)}\right)\cdot  \mathbf{n} \, \mathrm{d}S  \nn \\ 
&& \qquad  +
 \int_V\left( \tilde{\boldsymbol{v}}_f^{(2)}\cdot \tilde{\mathbf{f}}^{(1)}_f+ \tilde{\boldsymbol{v}}_n^{(2)}\cdot \tilde{\mathbf{f}}^{(1)}_n\right) \mathrm{d}V.
 \label{eq:RT2} 
\end{eqnarray}
Eq. \eqref{eq:RT2} is the reciprocal theorem for poro-viscoleastic (PVE) materials we set out to derive, and the main result of this work. Note that we have made no assumption about material isotropy up to this point, and the derived reciprocal relationship also applies to linear anisotropic materials. 
For a linear isotropic material, the network stress in Laplace space simplifies to $\tilde{\boldsymbol{\sigma}}_n(s)=\tilde{G}(s)\left(\nabla \tilde{\boldsymbol{v}}_n+\nabla \tilde{\boldsymbol{v}}^T_n\right)+\tilde{\lambda}(s)\left(\nabla \cdot \tilde{\boldsymbol{v}}_n\right)$,  
where, $\tilde{G}(s)$ and $\tilde{\lambda}(s)$ are the Laplace transforms of the time-dependent first and second Lam\'e coefficients. For a linear elastic network (no viscous component) these coefficients are related through the Poisson ratio, $\nu$: $\tilde{\lambda}=\frac{2\tilde{G}\nu}{1-2\nu}$. 
Hereafter, we omit $\sim$ sign for simplicity, noting the fact that the relevant quantities are in Laplace space.
\begin{figure}
    \centering
    \includegraphics[width=0.6\textwidth , height=5.5cm ]{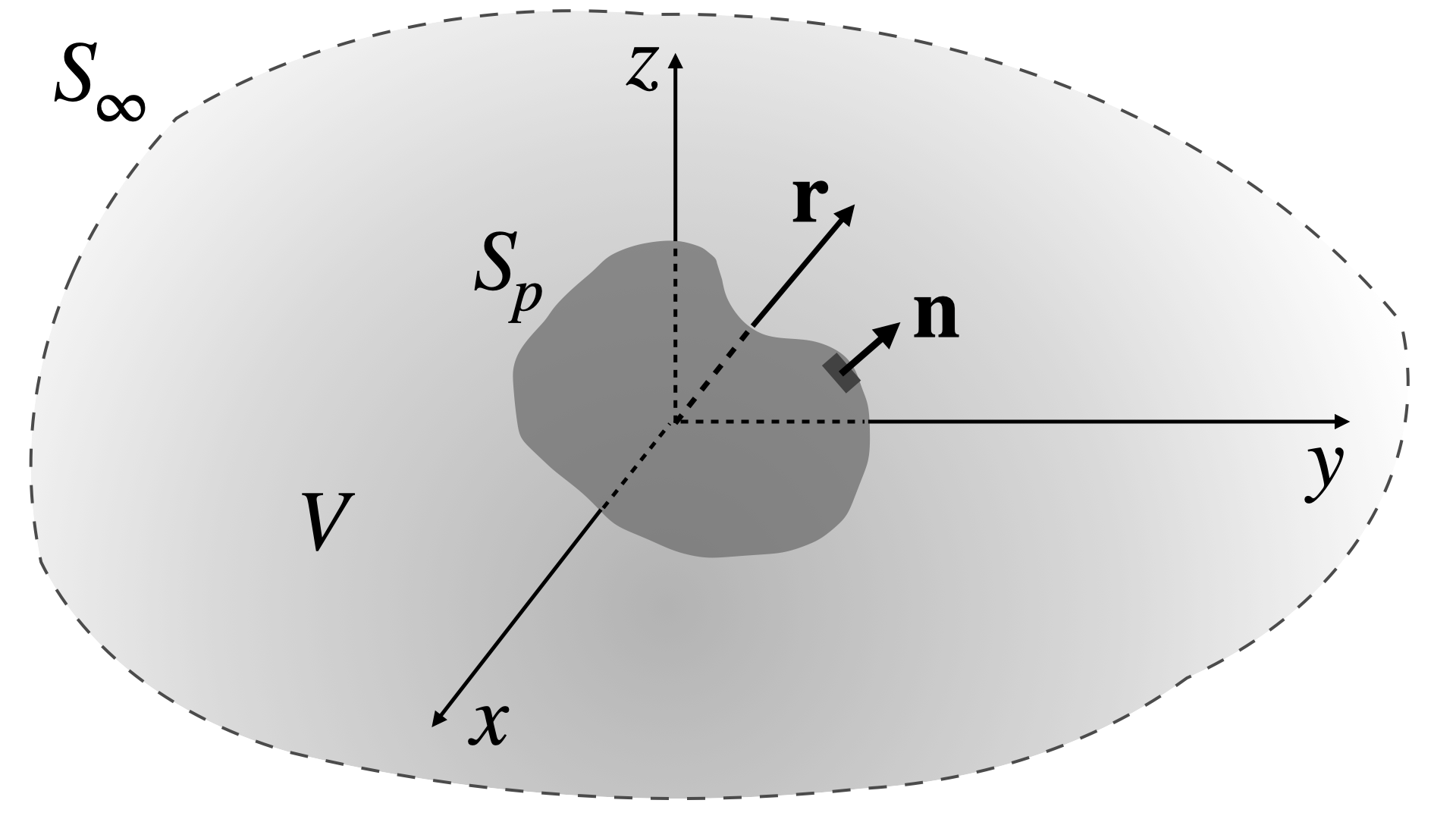} \label{fig1}
    \caption{ A schematic of an arbitrarily shaped inclusion, with surface $S_p$ and unit outward normal vector $\mathbf{n}$, in an unbounded domain. The dashed line indicates an enclosing boundary in the “far field”.
     }
    \label{fig1}
\end{figure}

\section{ force on a rigid sphere due to point-forces in fluid and network phases}
\label{sec:results}
\begin{figure}
    \centering 
    \begin{subfigure}[t]{0.48\textwidth}
         \centering
         \includegraphics[width=\textwidth  , height=6.3cm]{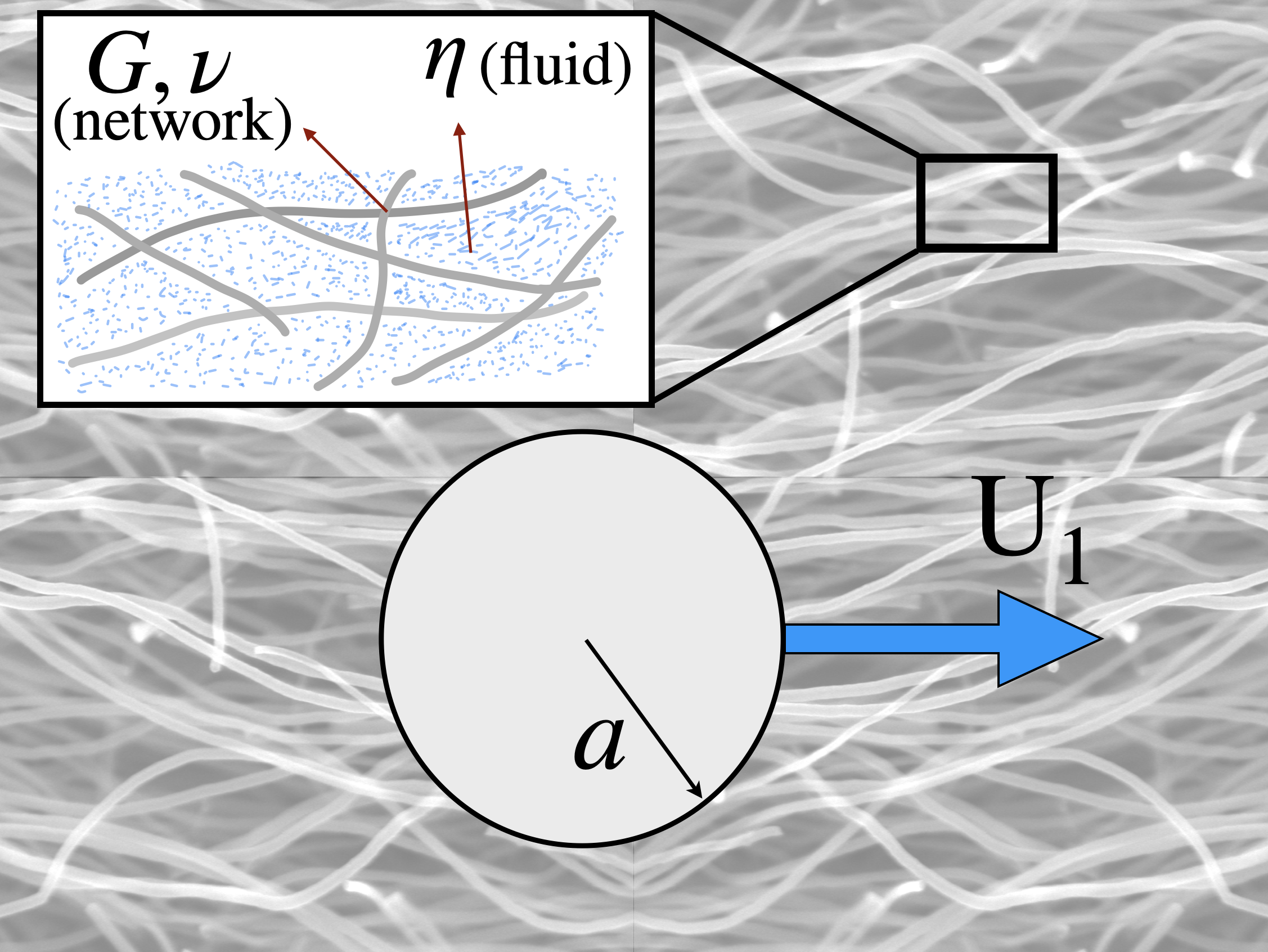}
         \caption{}
     \end{subfigure}
     \hfill
     \begin{subfigure}[t]{0.48\textwidth}
         \centering
         \includegraphics[width=\textwidth  , height=6.3cm]{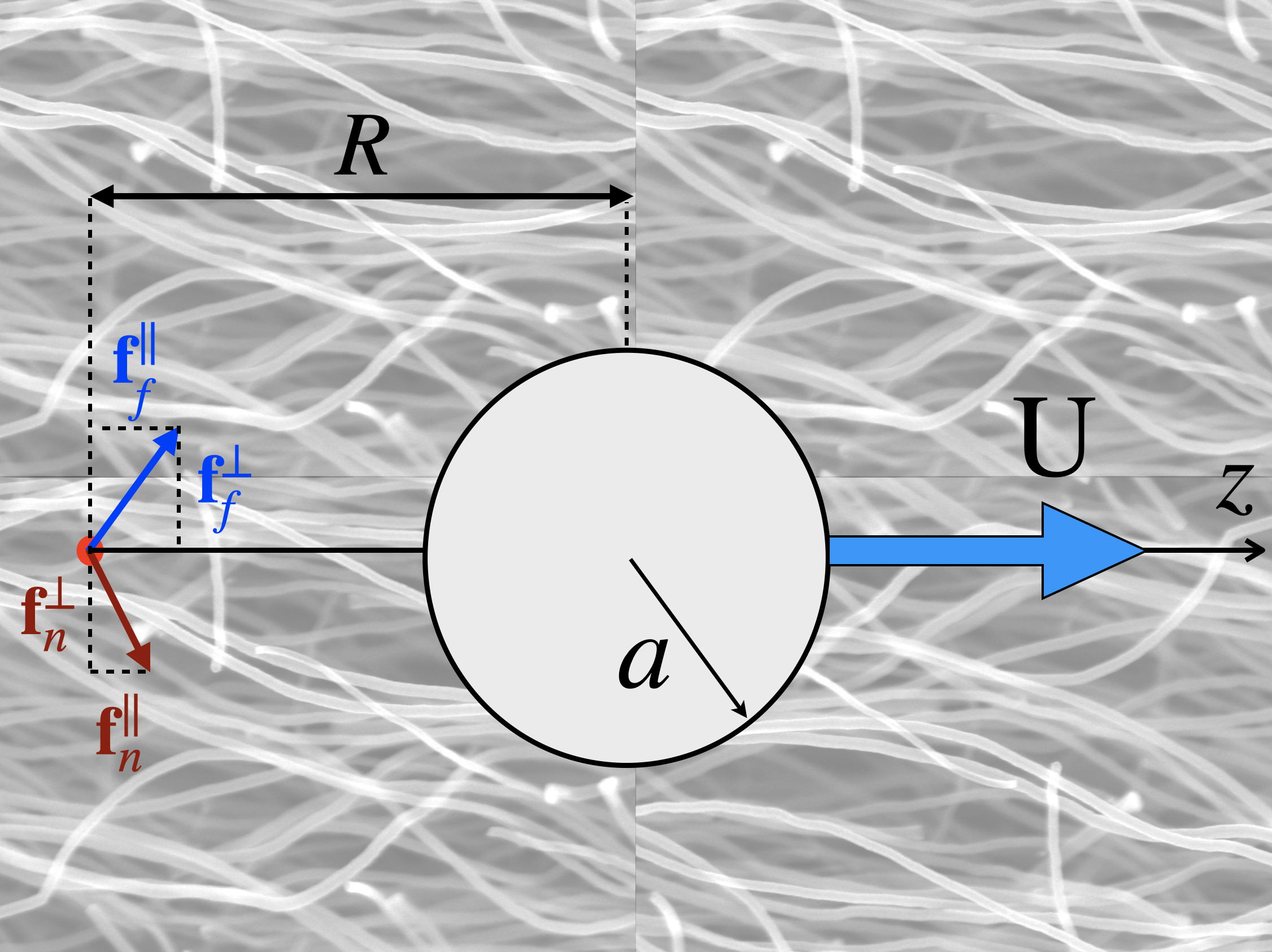}
         \caption{}
     \end{subfigure}
     \caption{ (a) \emph{The auxiliary problem in the implementation of the reciprocal theorem:} A rigid sphere moving with a prescribed velocity $\mathbf{U}_1$ in a linear PVE medium with fluid shear viscosity $\eta$ and network shear modulus $G$ and Poison ratio $\nu$. (b) \emph{The problem to be solved using reciprocal theorem:}  a rigid sphere moving with a prescribed velocity and subjected to external point-forces in the fluid and the network phases, ${\mathbf{f}}_f$, and ${\mathbf{f}}_n$.} 
    \label{fig2}
\end{figure}

One application of reciprocal theorem is finding analytical (or pseudo-analytical) solutions in simple geometries, including spheres. Here, we consider a rigid spherical inclusion moving with a prescribed velocity $\mathbf{U}$ in a PVE medium. Our goal to derive a closed-form expression for the net force, $\mathbf{F}$, on the sphere in response to point-forces in fluid and network phases, $\mathbf{f}_{\{f,n\}}$ at some separation distance, $\mathbf{R}$, from the center of the sphere; see Fig. \ref{fig2}(a) for  a schematic of this problem. 
The solution to this problem can used in different applications. For example, in Stokes flow the disturbance flow of a particle can be computed using  \emph{singularity solutions} technique, which represents the particle with a distribution of point-force, stresslets and higher moments of force \citep{kim2013microhydrodynamics}. 

We assume the network is a linear elastic material with shear modulus $G$ and Poisson ratio $\nu$, and the permeating fluid is Newtonian with shear viscosity $\eta$, so that $\tau (s)=s \eta /G=s\tau_\circ$, where $\tau_\circ=\eta/G$ is the shear relaxation time of the network in the fluid phase. However, the expressions that we are about to obtain can easily be extended to other linear viscoelastic description of fluid and network phases by modifying the expression for $\tau(s)$ in terms of constitutive parameters of the two phases.  
The flow fields and the net forces on the particle are first computed in Laplace space. We use Fourier–Euler summation (\citep{abate1992fourier}) to numerically invert  the results from $s$-space to time-space.

To use the reciprocal theorem we need a simple auxiliary problem, which gives us $\mathbf{v}_{\{f,n\}}^{(1)}$ and $\mathbf{\Sigma}_{\{f,n\}}^{(1)}$. 
We choose a sphere translating with velocity $\mathbf{U}_1$ as our auxiliary problem. We have obtained these solutions in our earlier work \citep{moradi2022general}. 
Since the general solution is axisymmetric (see Appendix \ref{App0}), it can be expressed in the following general form:
\begin{eqnarray}
&& {\boldsymbol{v}}_{f}^{(1)}  = \mathbf{U}_1 \cdot \Big[  {A}_{f} \, \hat{\boldsymbol{r}} \hat{\boldsymbol{r}} +  {B}_{f} \big(  \mathbf{I} - \hat{\boldsymbol{r}} \hat{\boldsymbol{r}} \big)      \Big] ,    \qquad\qquad    {\boldsymbol{v}}_{n}^{(1)}  = \mathbf{U}_1 \cdot \Big[  {A}_{n} \, \hat{\boldsymbol{r}} \hat{\boldsymbol{r}}   +  {B}_{n} \big(  \mathbf{I} - \hat{\boldsymbol{r}} \hat{\boldsymbol{r}} \big)  \Big], \qquad\qquad p^{(1)}=A_p \mathbf{U}_1 \cdot  \hat{\boldsymbol{r}}. 
\label{eq:aux}
\end{eqnarray}
Applying the boundary conditions,  ${\boldsymbol{v}}_f  {\big|}_{r=a} =  \mathbf{U}_1$, and ${\boldsymbol{v}}_n  {\big|}_{r=a} =  \mathbf{U}_1$, specifies, $A_{\{f,n\}}$ and $B_{\{f,n\}}$ as follows: 
\begin{subequations}
\begin{eqnarray}
&& A_f = \frac{1}{\Delta} \Big[   S_1 \big( \frac{a}{r} \big) + S_2 {\big( \frac{a}{r} \big)}^3 -3 \frac{a {\alpha}^2}{{\beta}^4 r^3} (1+ \beta r)  \, e^{-\beta (r-a)} \Big], \\
&& B_f = \frac{1}{\Delta} \Big[   S_1 \big( \frac{a}{r} \big) - S_2 {\big( \frac{a}{r} \big)}^3 +3 \frac{a {\alpha}^2}{{\beta}^4 r^3} (1+ \beta r + {\beta}^2 r^2 )  \, e^{-\beta (r-a)} \Big], \\
&&  A_n = \frac{1}{\Delta} \Big[   S_1 \big( \frac{a}{r} \big) - S_3 {\big( \frac{a}{r} \big)}^3 + 3 \tau \frac{a {\alpha}^2}{{\beta}^4 r^3} (1+ \beta r)  \, e^{-\beta (r-a)}  + 3 \frac{a (1+\tau)}{{\beta}^2 r^3 } (2+2\alpha r + {\alpha}^2 r^2 ) \, e^{-\alpha (r-a)} \Big],  \\
&& B_n = \frac{1}{\Delta} \Big[   S_1 \big( \frac{a}{r} \big) + S_3 {\big( \frac{a}{r} \big)}^3 -3 \tau \frac{a {\alpha}^2}{{\beta}^4 r^3} (1+ \beta r + {\beta}^2 r^2 )  \, e^{-\beta (r-a)} -6 \frac{a (1+\tau)}{{\beta}^2 r^3 } (1+\alpha r  ) \, e^{-\alpha (r-a)}\Big] , \\
&& A_p = \frac{G}{a \Delta} \Big[ \frac{1+\tau}{3} S_1  {\big( \frac{a}{r} \big)}^2 - 3 \tau {\big( \frac{a}{r} \big)}^2   (1+ \alpha r)  \, e^{-\alpha (r-a)}   \Big] ,
\end{eqnarray}
\label{eq:AfBf}
\end{subequations}
where
\begin{align}
   S_1 &= 3 \tau \big( 1+ a \alpha + \frac{1}{2} \frac{{\alpha}^2}{{\beta}^2} (1+ a \beta + a^2 {\beta}^2)  \big),&  S_2 &= \frac{{\alpha}^2}{a^2 {\beta}^4} (3+ 3a \beta + a^2 {\beta}^2 ) - \frac{1}{3} S_1,&  \\ \nonumber S_3 &= \frac{2}{a^2 {\beta}^2} (3+3a \alpha + a^2 {\alpha}^2) + \frac{6+ a^2 {\beta}^2}{3 a^2 {\beta}^2} S_1,& \Delta &= \frac{{\alpha}^2}{{\beta}^2} + \frac{2}{3} S_1  .& 
\end{align}
Here, $\tau (s) =s \frac{\eta}{G}=s\tau_\circ$ is the shear relaxation time, and ${\beta}^2 = {\beta}_{\circ}^2 (1+ \tau  )$, with ${\beta}_{\circ} =\sqrt{ \frac{\xi}{\eta}}$ being the inverse of permeability of the fluid  and ${\alpha}^2 = s \tau_\circ \alpha_\circ^2$, where  $\alpha_\circ^2={\beta}_{\circ}^2   \frac{ 1-2\nu }{2(1-\nu)}$ with $\nu$ being the Poisson ratio. 
For an incompressible network, $\nu=0.5$ and $\alpha_\circ=0$. Substituting these values in the above equations cancels all the terms involving $\beta_\circ$ and reduces the expressions for $A_{\{f,n\}}$ and $A_{\{f,n\}}$  in Eq. \eqref{eq:AfBf}, to their well-known form in Stokes flow: $$ A_{\{f,n\}}(\nu=0.5)=A_\text{Stokes}=\frac{3}{2}\left(\frac{a}{R}\right)-\frac{1}{2}\left(\frac{a}{R}\right)^3,\quad B_{\{f,n\}}(\nu=0.5)=B_\text{Stokes}=\frac{3}{4}\left(\frac{a}{R}\right)+\frac{1}{4}\left(\frac{a}{R}\right)^3. $$

We are interested in the dynamical features that arise due to network compressibility. 
Using the expressions in Eq. \eqref{eq:aux}, we can compute the traction applied by the fluid and the network phase on the surface of the sphere and integrate those on the sphere's surface to get the net force on the particle:
\begin{equation}
\,\, \boldsymbol{F}_1 =   \int  \big( {\boldsymbol{\Sigma}}_f^{(1)}  + {\boldsymbol{\Sigma}}_n^{(1)}  \big) \cdot \mathbf{n}\, \mathrm{d}S \equiv - \mathcal{R} \mathbf{U}_1 , \qquad 
\mathcal{R} =  6 \pi \eta  a \Bigg[ \frac{ ( 1  + \tau   ) \left(\alpha ^2 \left(a^2 \beta ^2+a \beta +1\right)+   2  \beta ^2 (1+a\alpha)   \right)}{ \tau   \big( \alpha ^2 \left(a^2 \beta ^2+a \beta +1\right)+2  \beta ^2 (1+a\alpha) \big)+\alpha ^2   }   \Bigg] . \label{FUrelation}
\end{equation}
The term between brackets is the correction to Stokes drag due to network compressibility. 

The second problem is the problem we seek to solve, which is shown in Fig. \ref{fig2}(b). We assume no-slip BCs for both the fluid and network on the surface of the sphere: $ {\boldsymbol{v}}_{\{f , n \}}^{(2)}(r=a) = \mathbf{U}$. No-slip BCs is generally valid for the fluid phase. The BCs for the network phase is more complex \citep{fu2010low}. When the particle size is much larger than the network's mesh size, $a\beta_\circ \gg 1$, the no-slip BC is a good approximation. This is the BC used in this study.  
The point-forces are expressed as ${\mathbf{f}}_f^{(2)} = {\mathbf{f}}_{f} \delta (\boldsymbol{r}- \boldsymbol{R})$ and ${\mathbf{f}}_n^{(2)} = {\mathbf{f}}_{n} \delta (\boldsymbol{r}- \boldsymbol{R})$, where $\delta (\boldsymbol{r}- \boldsymbol{R})$ is the Dirac delta function. After substituting these expressions in Eq. \eqref{eq:RT2} and a few simple lines of algebra we get
\begin{eqnarray}
&& - \mathcal{R}  \mathbf{U} +  \boldsymbol{\mathcal{M}}_{f} \cdot \mathbf{f}_f +  \boldsymbol{\mathcal{M}}_{n} \cdot \mathbf{f}_n = \mathbf{F},
\label{eq:F1}
\end{eqnarray}
\begin{align}
& \text{where}&  & \boldsymbol{\mathcal{M}}_{f} =   A_f  (R) \hat{\boldsymbol{R}} \hat{\boldsymbol{R}} +  B_f (R) \big(  \mathbf{I} - \hat{\boldsymbol{R}} \hat{\boldsymbol{R}} \big), & & 
\boldsymbol{\mathcal{M}}_{n}  =  A_n (R) \hat{\boldsymbol{R}} \hat{\boldsymbol{R}}   +  B_n (R) \big(  \mathbf{I} - \hat{\boldsymbol{R}} \hat{\boldsymbol{R}} \big), &
\label{eq:F2}
\end{align}
and $\hat{\mathbf{R}}=\mathbf{R}/|\mathbf{R}|$. Equations \eqref{eq:F1} and \eqref{eq:F2} are the expressions we set out to derive. $\boldsymbol{\mathcal{M}}_{f}$ and  $\boldsymbol{\mathcal{M}}_{n}$ are the response functions that relate the forces on  fluid and network phases to the net force and velocity of the spherical particle. For a force-free particle (mobility problem), we can set $\mathbf{F}=0$, which yields 
\begin{align}
    &\text{Force-free particle,} \, \mathbf{F}=\mathbf{0}:& \mathbf{U}=\mathcal{R}^{-1}\left(\boldsymbol{\mathcal{M}}_{f} \cdot \mathbf{f}_f +  \boldsymbol{\mathcal{M}}_{n} \cdot \mathbf{f}_n\right),
    \label{eq:resistance}
\end{align}
In this setup $\boldsymbol{\mathcal{M}}_{f}$ and $\boldsymbol{\mathcal{M}}_{n}$ can be thought of as mobility tensors that relate the forces on each phase to the particle's velocity.  For brevity we only present the results of 
fixed particle (resistance problem), $\mathbf{U}=\mathbf{0}$, in the next section. The solution can easily be extended to force-free particles (mobility problem), using Eq. \eqref{eq:resistance}.

\subsection*{Net force on a fixed sphere}
We consider the special case of a fixed particle, $\mathbf{U} =\mathbf{0}$, which simplifies Eq. \eqref{eq:F1} to 
\begin{align}
    &\text{Fixed particle,} \, \mathbf{U}=\mathbf{0}:& \mathbf{F}=\boldsymbol{\mathcal{M}}_{f} \cdot \mathbf{f}_f +  \boldsymbol{\mathcal{M}}_{n} \cdot \mathbf{f}_n.
\end{align}
It is useful to decompose the point-forces and the force on the particle in parallel and perpendicular directions of the separation vector.
Using $\mathbf{f}_{\{f,n\}} = \mathrm{f}_{\{f,n\}}^{\parallel} \hat{\mathbf{R}} + \mathrm{f}_{\{f,n\}}^{\perp} \hat{\mathbf{R}}^{\perp}$, and $\mathbf{F}=F^\parallel \hat{\mathbf{R}}+F^\perp  \hat{\mathbf{R}}^{\perp}$ we get: 
\begin{align}
&  {F}^{\parallel} =   A_f (R) \mathrm{f}_f^{\parallel}  + A_n (R) \mathrm{f}_n^{\parallel}  ,&
&  {F}^{\perp} =    B_f (R) \mathrm{f}_f^{\perp} +  B_n (R) \mathrm{f}_n^{\perp}  .& 
\end{align}

We begin by analyzing the limiting behaviors of  $A_{\{f,n\}}$ and $B_{\{f,n\}}$ at early ($t\to 0$) and long ($t\to\infty$) times. These limits can be computed analytically using the following relationships:   $f(t\to 0^{+})= \lim_{s\to\infty} s f(s) $ and $f(t\to\infty)= \lim_{s\to 0} s f(s)$ and the results are:
\begin{subequations}
\begin{eqnarray}
&&   {A}_{ \{f , n\} } (t\to 0)=  \frac{3}{2} \big( \frac{a}{R} \big) -\frac{1}{2} {\big( \frac{a}{R} \big)}^3  , \qquad\qquad   {B}_{ \{f , n\} } (t\to 0)=  \frac{3}{4} \big( \frac{a}{R} \big) +  \frac{1}{4} {\big( \frac{a}{R} \big)}^3  ,   \\
&&  A_f (t\to \infty) =  \frac{6(1-\nu)}{5-6\nu} \big( \frac{a}{R} \big) - \frac{1}{5-6\nu}  {\big( \frac{a}{R} \big)}^3  + \frac{3a (1-2\nu)}{{\beta}_{\circ}^2 R^3 (5-6\nu)} \Big[   1+a {\beta}_{\circ}  -(1+ {\beta}_{\circ} R  )  {\mathrm{e}}^{-{\beta}_{\circ} (R-a)}  \Big]  ,  \\
&& B_f (t\to \infty) =  \frac{3(1-\nu)}{5-6\nu} \big( \frac{a}{R} \big) + \frac{1}{2(5-6\nu)}  {\big( \frac{a}{R} \big)}^3  - \frac{3a (1-2\nu)}{2 {\beta}_{\circ}^2 R^3 (5-6\nu)} \Big[   1+a {\beta}_{\circ}  -(1+ {\beta}_{\circ} R + {\beta}_{\circ}^2 R^2   )  {\mathrm{e}}^{-{\beta}_{\circ} (R-a)}  \Big]  , \\
&&  A_n (t\to \infty) = \frac{6(1-\nu)}{5-6\nu} \big( \frac{a}{R} \big) - \frac{1}{5-6\nu}  {\big( \frac{a}{R} \big)}^3   , \\
&&  B_n (t\to \infty) =  \frac{3(3-4\nu)}{2(5-6\nu)} \big( \frac{a}{R} \big) + \frac{1}{2(5-6\nu)}  {\big( \frac{a}{R} \big)}^3. 
\end{eqnarray}
\label{eq:limits}
\end{subequations}

As shown in Eq. \eqref{eq:limits}(a) at short times the response functions for both fluid and network asymptote to their well-known respective forms in Stokes flow \citep{kim2013microhydrodynamics}. 
This can be explained by noting that at early times the network co-moves with the fluid phase, $\mathbf{v}_n(\mathbf{x},t\to 0)=\mathbf{v}_f(\mathbf{x},t\to 0)$, and $\nabla \cdot \mathbf{v}_{\{f,n\}}(\mathbf{x},t\to 0)=0$. Furthermore, at long times the response functions of the network phase ($A_n$ and $B_n$), approach their well-known form in linear elasticity \citep{phan1994microstructures}; see Eqs. \eqref{eq:limits}(d-e).    
The long time forms of the fluid response functions ($A_f$ and $B_f$) include extra $\beta_\circ$-dependent, which, as expected, become identically zero for an incompressible network. 

Variations of the net force with time and the distance from the point-forces, $\mathbf{F}(\boldsymbol{R},t)$, are determined by the behavior of $A_{\{f,n\}}$. Since we have the analytical form of these functions in short and long times in Eq. \eqref{eq:limits}, we study the time-dependent behavior of the following quantities
\begin{align}
    \hat{A}_{\{f,n\}}&= \frac{A_{\{f,n\}}(t,R)-A_{\{f,n\}}(t\to \infty,R)}{A_{\{f,n\}}(0,R)-A_{\{f,n\}}(t\to \infty,R)},& \hat{B}_{\{f,n\}}&=\frac{B_{\{f,n\}}(t,R)-B_{\{f,n\}}(t\to \infty,R)}{B_{\{f,n\}}(0,R)-B_{\{f,n\}}(t\to \infty,R)}.&
    \label{eq:Afn}
\end{align}
$\hat{A}$ and $\hat{B}$, by construction relax over time from $\hat{A}=\hat{B}=1$ at $t=0$ to $\hat{A}=\hat{B}\to 0$ as $t\to \infty$. 
We would like to explore how this relaxation dynamics changes with the separation distance, $R$, and material properties of the fluid and network i.e. $\eta, G, \, \beta_\circ$ and $\nu$. 

The expressions for $A_{\{f,n\}}$ and $B_{\{f,n\}}$ in Eqs. \eqref{eq:AfBf} contain two types of $s$ and $R$ dependencies. One obvious relaxation mechanism is shear relaxation, $\tau_\circ=\eta/G$, which is independent of $\beta_\circ$ and $\nu$ and the separation distance, $R$. This is the relaxation time measured in the standard particle tracking microrheology.  
The terms involving $S_1$, $S_2$ and $S_3$ can be written as a product of an $s-$dependent and an $r-$dependent function, $\tilde{\mathcal{Q}}(s) \mathcal{P}(r)$, resulting in $\mathcal{Q}(t) \mathcal{P}(r)$ form after Laplace inversion. As a result, the relaxation time becomes independent of $R$. In comparison, the terms that include $\exp(-\beta (s) r)$ and $\exp(-\alpha (s) r)$, cannot be written as a product of $s-$ and $r-$dependent functions, and so the relaxation time of the force on the sphere is a function of the distance from the point-forces. 
\begin{figure}[!b]
    \centering 
    \begin{subfigure}[t]{0.3\textwidth}
         \centering
         \includegraphics[width=\textwidth]{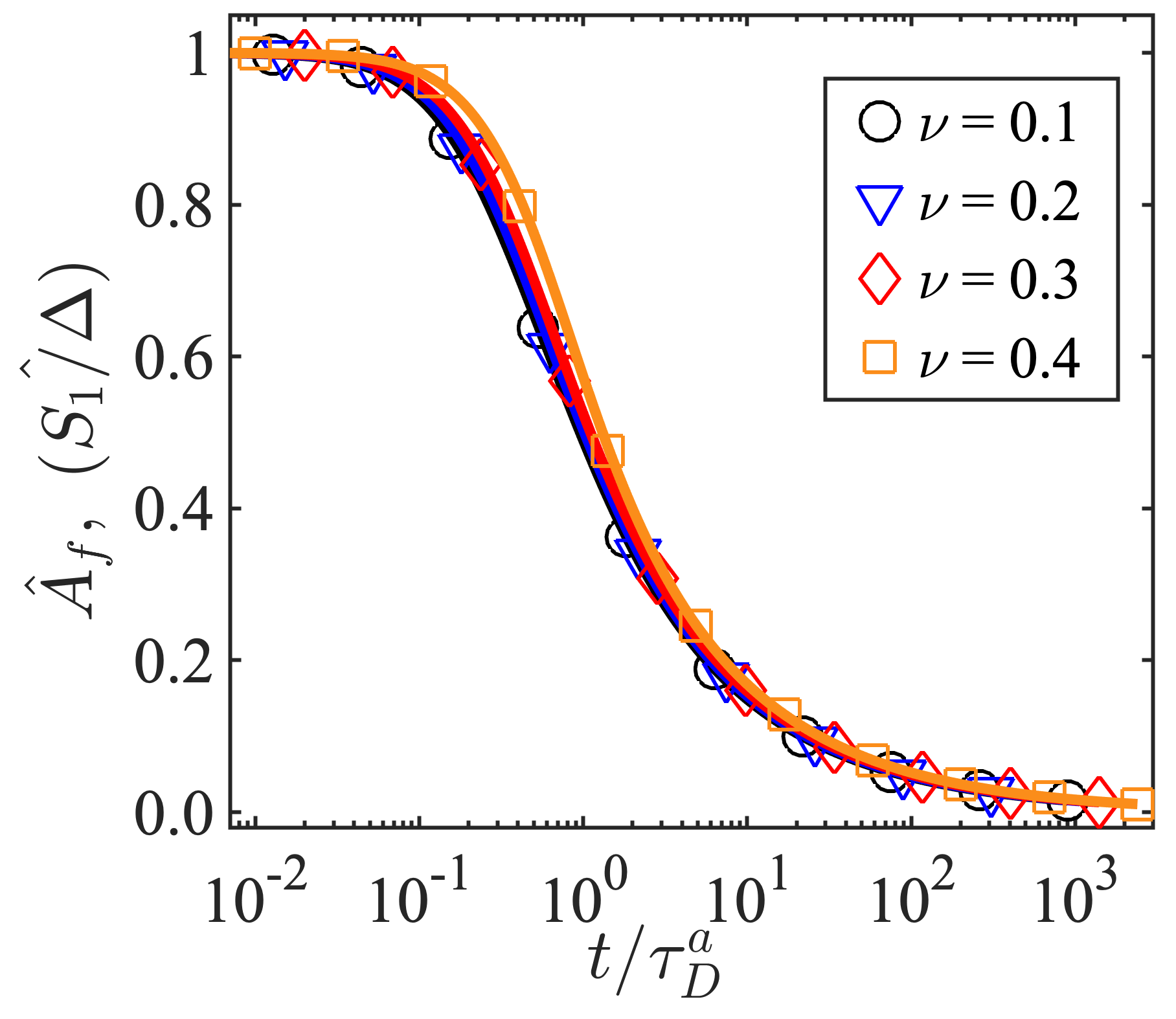}
         \caption*{}
     \end{subfigure}
     \hfill
     \begin{subfigure}[t]{0.3\textwidth}
         \centering
         \includegraphics[width=\textwidth]{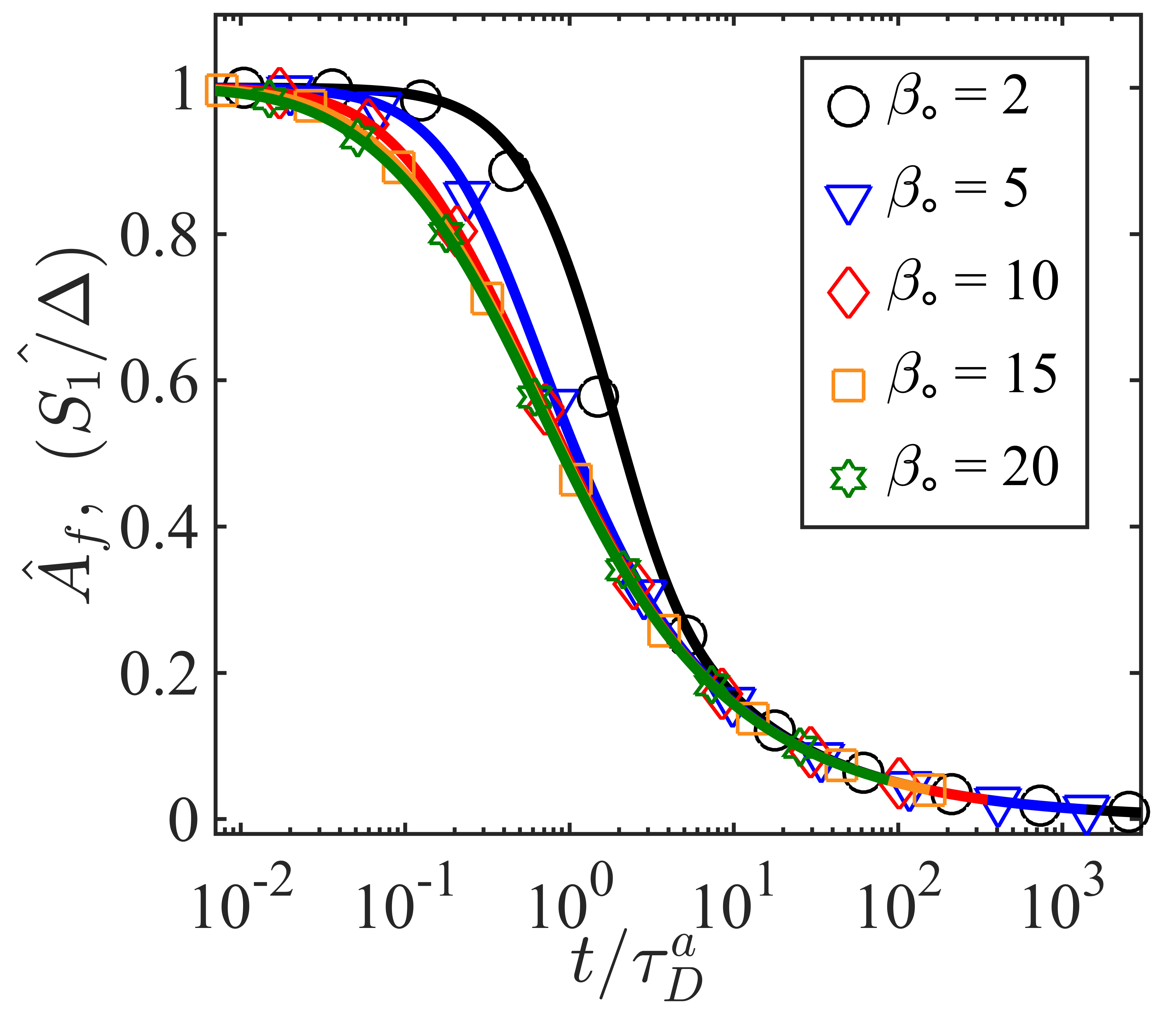}
         \caption*{}
     \end{subfigure}
     \hfill
     \begin{subfigure}[t]{0.3\textwidth}
         \centering
         \includegraphics[width=\textwidth]{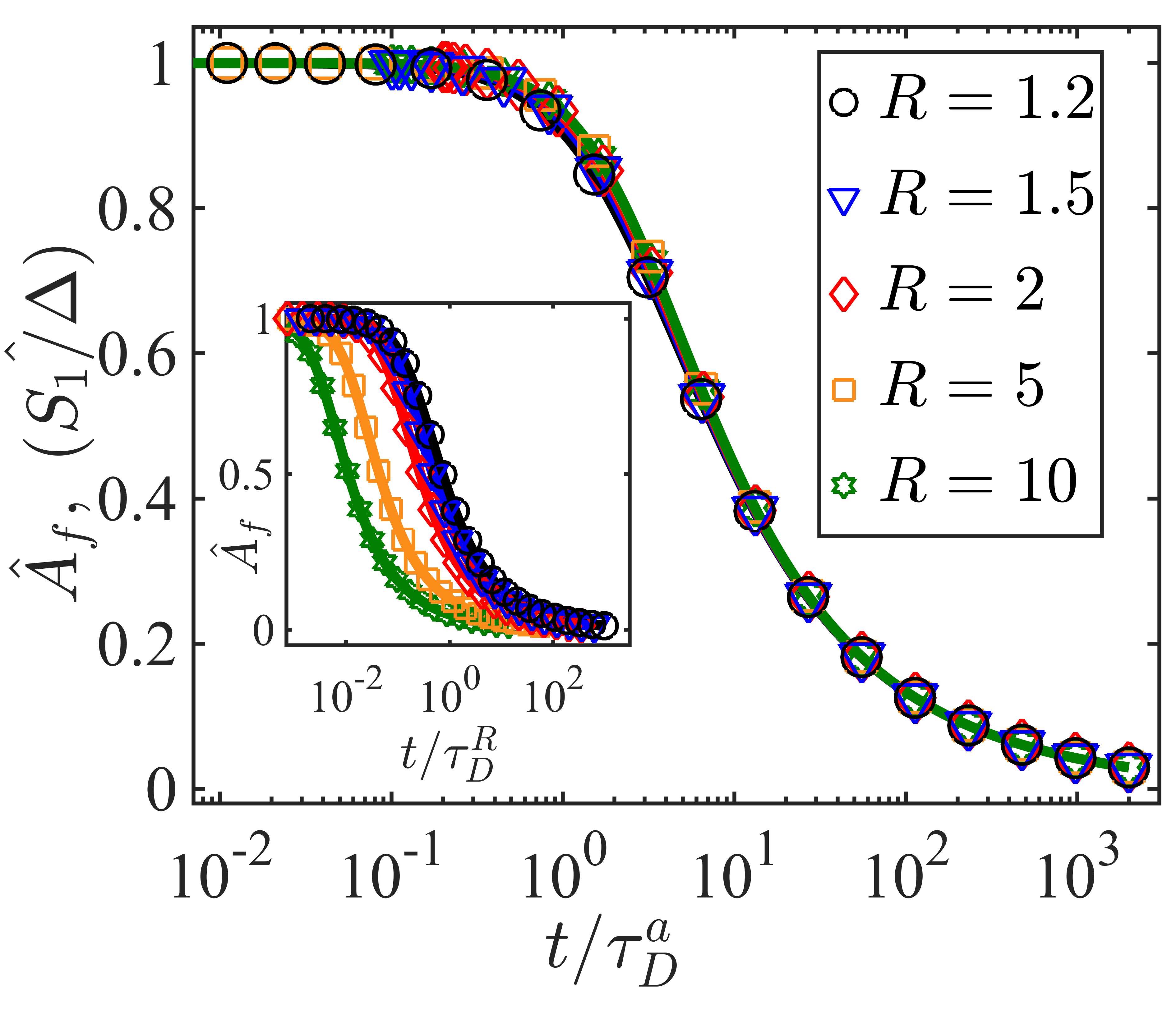}
         \caption*{}
     \end{subfigure}
     \\
     \begin{subfigure}[t]{0.3\textwidth}
         \centering
         \includegraphics[width=\textwidth]{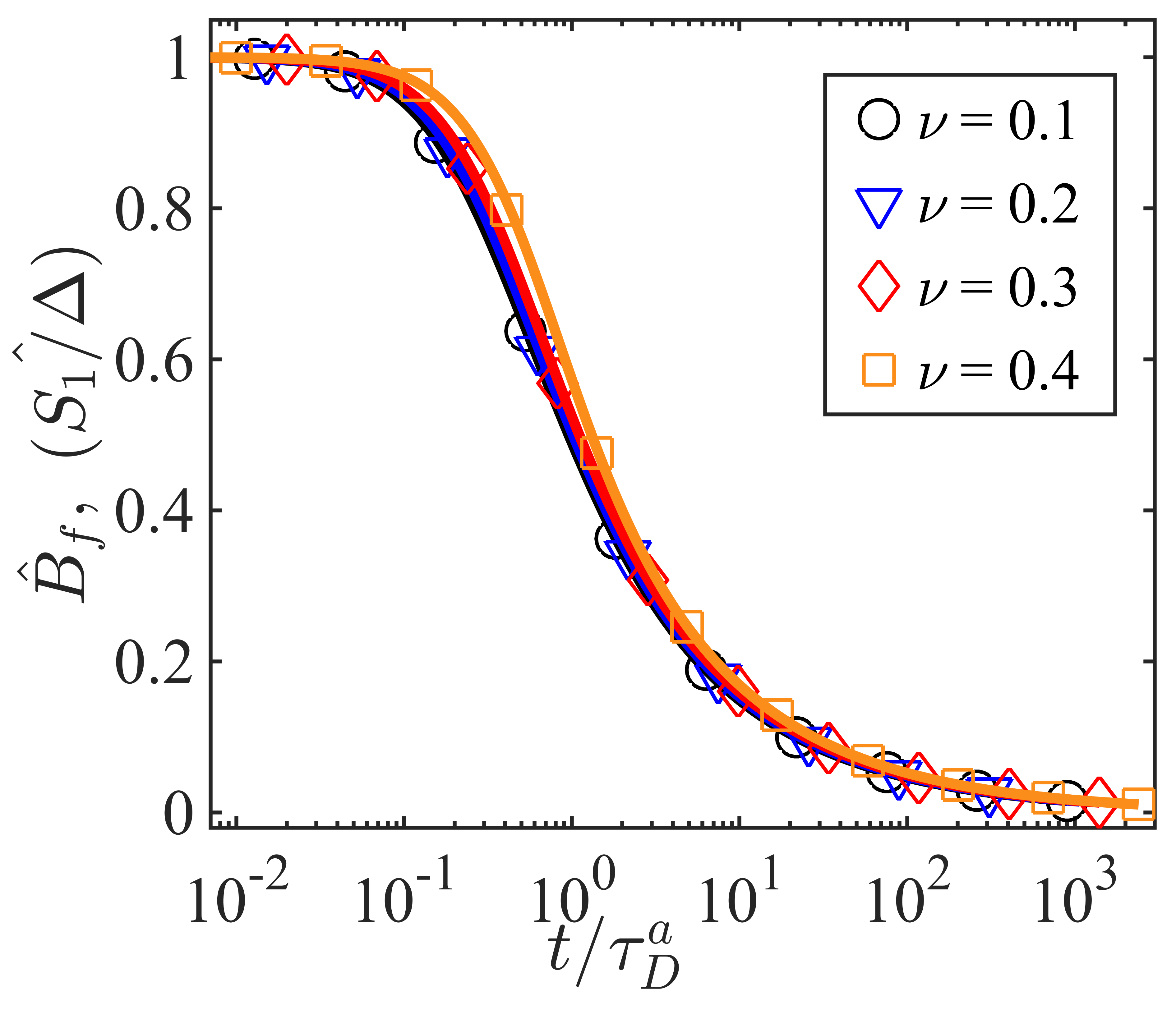}
         \caption{Varying Poisson ratio, $\nu$}
     \end{subfigure}
     \hfill
     \begin{subfigure}[t]{0.3\textwidth}
         \centering
         \includegraphics[width=\textwidth]{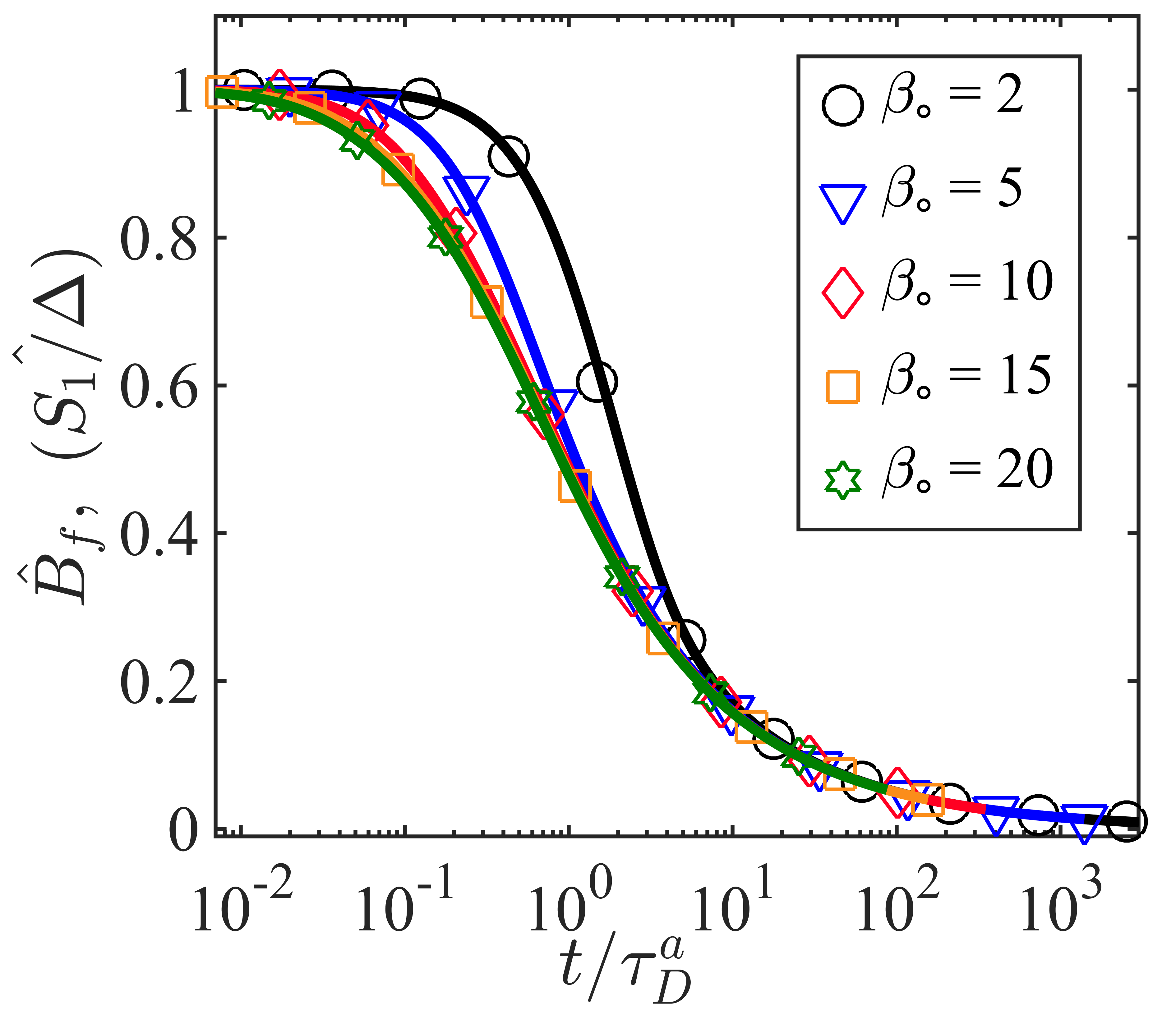}
         \caption{Varying Permeability, $\beta_\circ$}
     \end{subfigure}
     \hfill
     \begin{subfigure}[t]{0.3\textwidth}
         \centering
         \includegraphics[width=\textwidth]{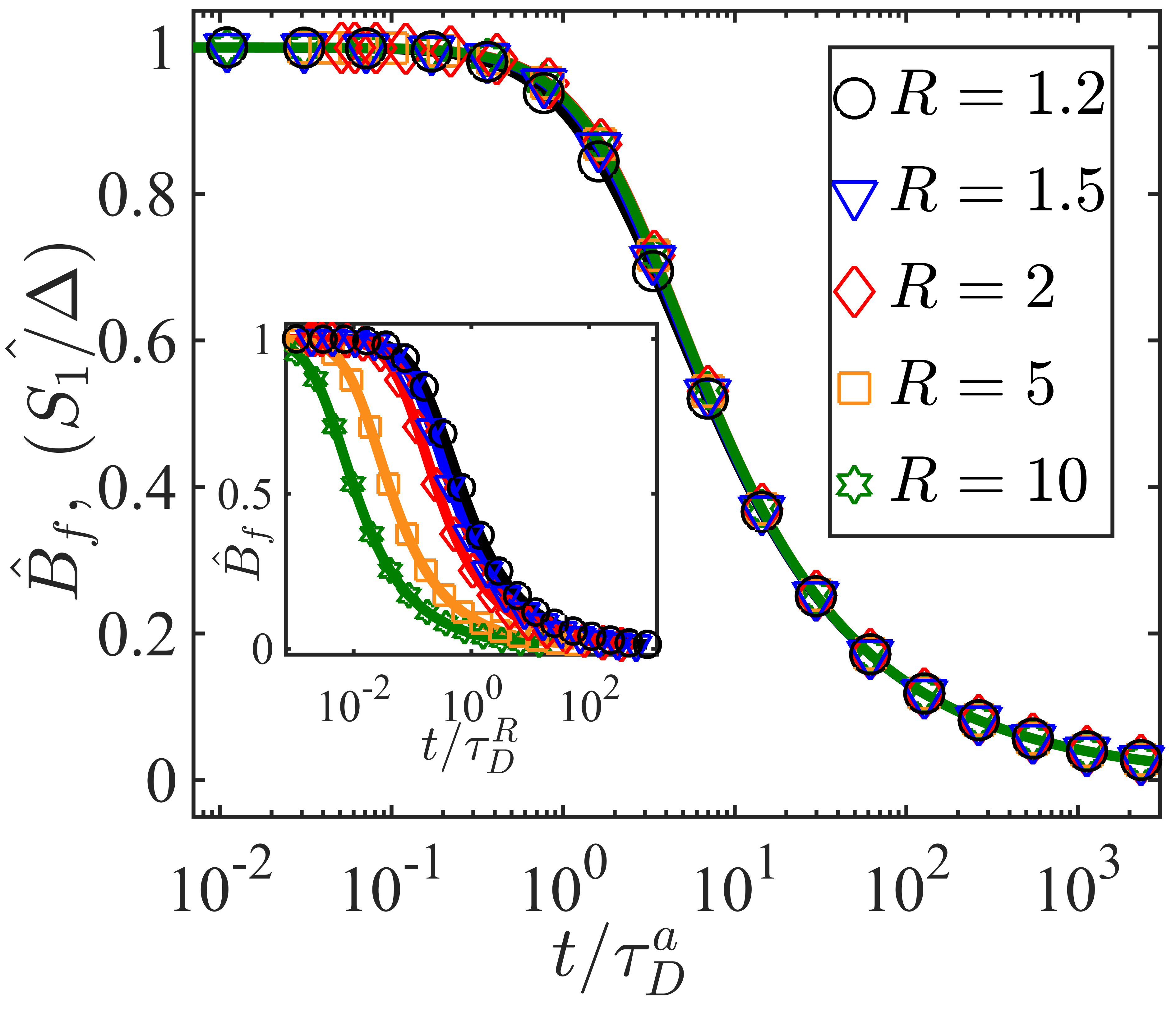}
         \caption{Varying $R$}
     \end{subfigure}
\caption{\emph{Time variations of the fluid hydrodynamic functions, $\hat{A}_f$ (top row) and $\hat{B}_f$ (bottom row).} (a) Different values of Poisson ratio, $\nu$, while setting $\beta_\circ=5$ and $R=2$. (b) Different values permeability, $\beta_\circ$, while setting $\nu=0.3$ and $R=2$, and (c) different separation distances, $R$, while setting $\nu=0.3$ and $\beta_\circ=5$. 
The time axis is made dimensionless with $\tau_D^a=a^2\alpha_\circ^2\tau_\circ$ in all the main plots. The values of $R$ and $\beta_\circ$ are made dimensionless with the sphere radius, which is equivalent to taking $a:=1$. 
The inset figures in Fig.~\ref{fig3}(c) show the same results as the main figure, but the time axes are made dimensionless with $\tau_D^R=R^2\alpha_\circ^2\tau_\circ$.}
        \label{fig3}
\end{figure}
As shown in our earlier work \citep{moradi_shi_nazockdast_2022} and others \citep{doi2009gel, detournay1993fundamentals} the divergence of the displacement field in a linear elastic network , $\nabla \cdot \mathbf{v}_n$, is described by a diffusion equation, where the diffusion coefficient is $D=\tau_\circ \alpha_\circ^{-2}$. 
Based on this, we can introduce two extra timescales: $\tau_D^a=(\alpha_\circ a)^2\tau_\circ$, which is the timescale for the network compressibility to diffuse over a particle radius, $a$; and $\tau_D^{R}=(\alpha_\circ R)^2\tau_\circ$ is the time for diffusing distance $R$ from the point-forces. 

If the $S_1$, $S_2$ and $S_3$ terms dominate the response, we expect the force relaxation time to be independent of $R$ and determined by $\tau_D^ a$; alternatively, if $\exp(-\alpha(s) r)$ and $\exp(-\beta(s) r)$ term determine the behavior of $A_{\{f,n\}}$ and $B_{\{f,n\}}$ we expect the force relaxation to be distance-dependent and in part determined by $\tau_D^{R}$. 

We begin with presenting the results for the relaxation dynamics of fluid phase functions, $\hat{A}_f$ and $\hat{B}_f$ vs time in Fig. \ref{fig3} for different values of Poisson ratio, $\nu$ (Fig \ref{fig3}(a)), permeability, $\beta_\circ$ (Fig \ref{fig3}(b)) and distance from the point-force $R$ (Fig \ref{fig3}(c)). The solid lines that go through the markers are the computed values of $\hat{\left(S_1/\Delta\right)}$, which is simply the the values of $S_1/\Delta$ made dimensionless in the exact manner as $A_f$ and $B_f$ in Eq. \eqref{eq:AfBf}.
The relaxation time is made dimensionless with $\tau_D^a$ in all figures. 
When $R$ is fixed and $\nu$ and $\beta_\circ$ are varied (Figs. \ref{fig3}(a-b)), $\tau_D^a$ and $\tau_D^R$ are linearly related by a constant factor, $\tau_D^R/\tau_D^a = (R/a)^2$, and using $\tau_D^R$ (instead of $\tau_D^a$) would only shift the plots in x-axis and would not change the overall superposition of the curves.
In comparison when $R$ is varied, the two timescales 
become distinct. Thus, to test if the relaxation is distance-dependent or not, we present the results of Figs \ref{fig3}(c)  as a function of $t/\tau_D^R$ in the insets of those figures.

As it can be seen, in all cases $\hat{\left(S_1/\Delta\right)}$ (solid lines) closely match the predictions that include all the terms (symbols). 
This suggest that \emph{the relaxation is distance-independent} and $\tau_D^a$ is a more appropriate choice for relaxation time of $\hat{A}_f$ and $\hat{B}_f$, compared to $\tau_D^R$. This can be clearly seen by comparing the nearly perfect superposition of the plots in the main figure against the inset plots in Fig. \ref{fig3}(c).  Furthermore, the relaxation curves for different values of $\nu$ superimpose, when time is made dimensionless with $\tau_D^a$. The superposition is not as clean at shorter times when permeability is varied (see Fig.~\ref{fig3}(b)), since shear relaxation time ($\tau_\circ$), becomes increasingly more important. However, at longer times ($t/\tau_D^a\ge 10$) the plots superimpose, showing that the long time behavior of these hydrodynamic functions is determined by $\tau_D^a$. 
Collectively, these results show that the relaxation dynamics of $\hat{A}_{f}$ and $\hat{B}_f$ is not distance-dependent, and is dominated by the terms involving the slowest decay with distance i.e. $(S_1/\Delta)(a/r)$ and the relaxation time, $\tau_D^a$. 
\begin{figure}[!b]
    \centering 
    \begin{subfigure}[t]{0.3\textwidth}
         \centering
         \includegraphics[width=\textwidth]{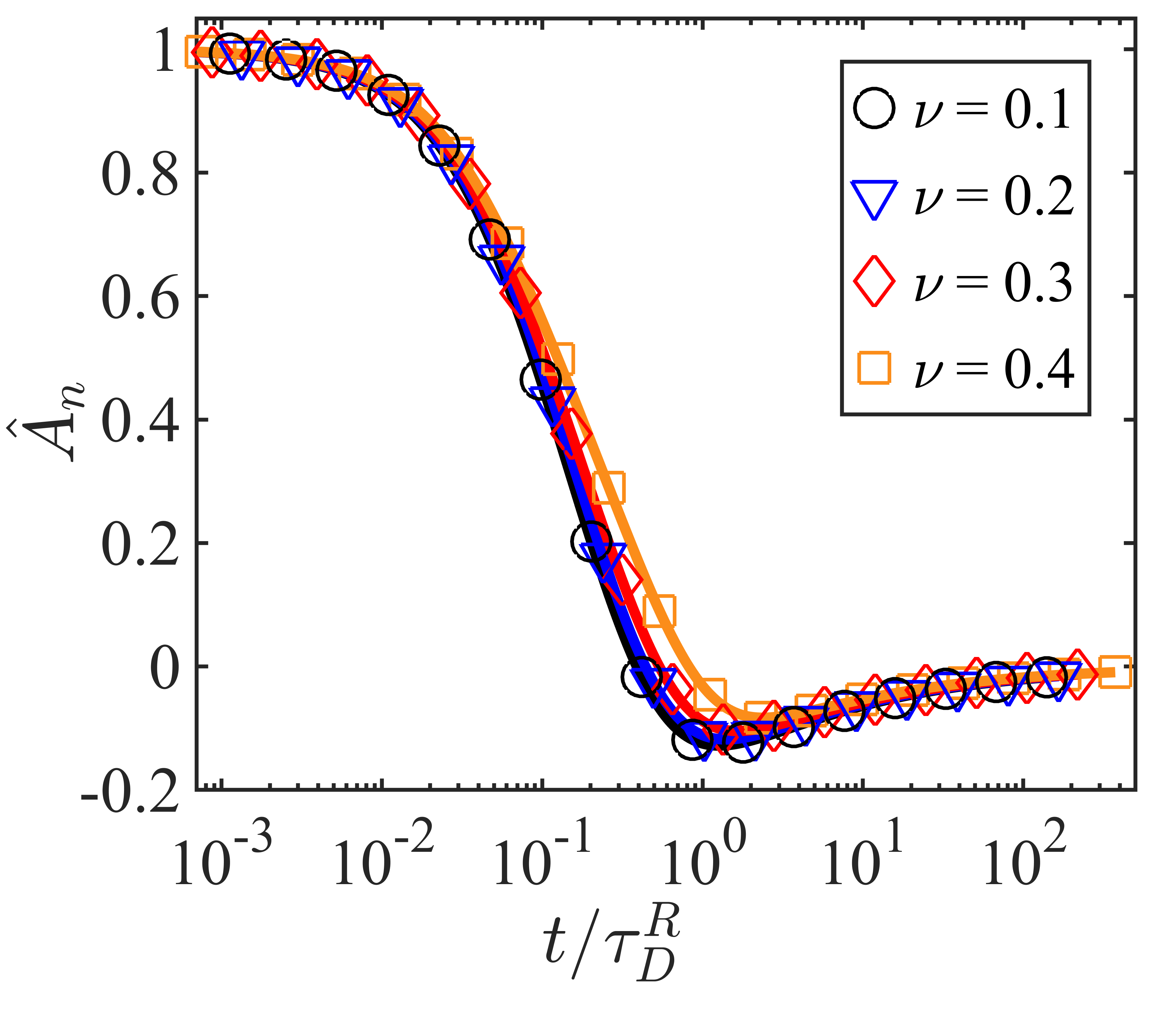}
         \caption*{}
     \end{subfigure}
     \hfill
     \begin{subfigure}[t]{0.3\textwidth}
         \centering
         \includegraphics[width=\textwidth]{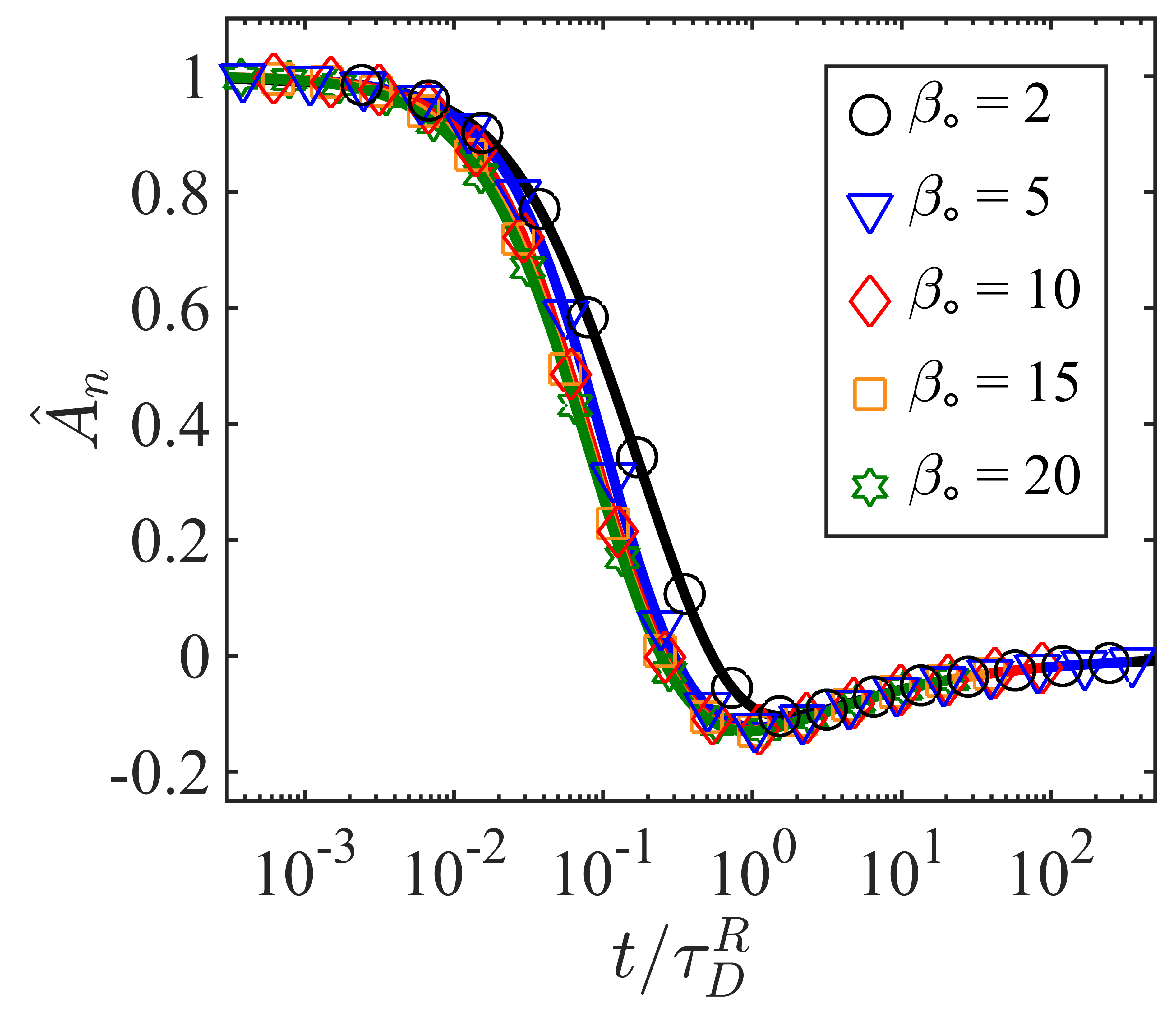}
         \caption*{}
     \end{subfigure}
     \hfill
     \begin{subfigure}[t]{0.3\textwidth}
         \centering
         \includegraphics[width=\textwidth]{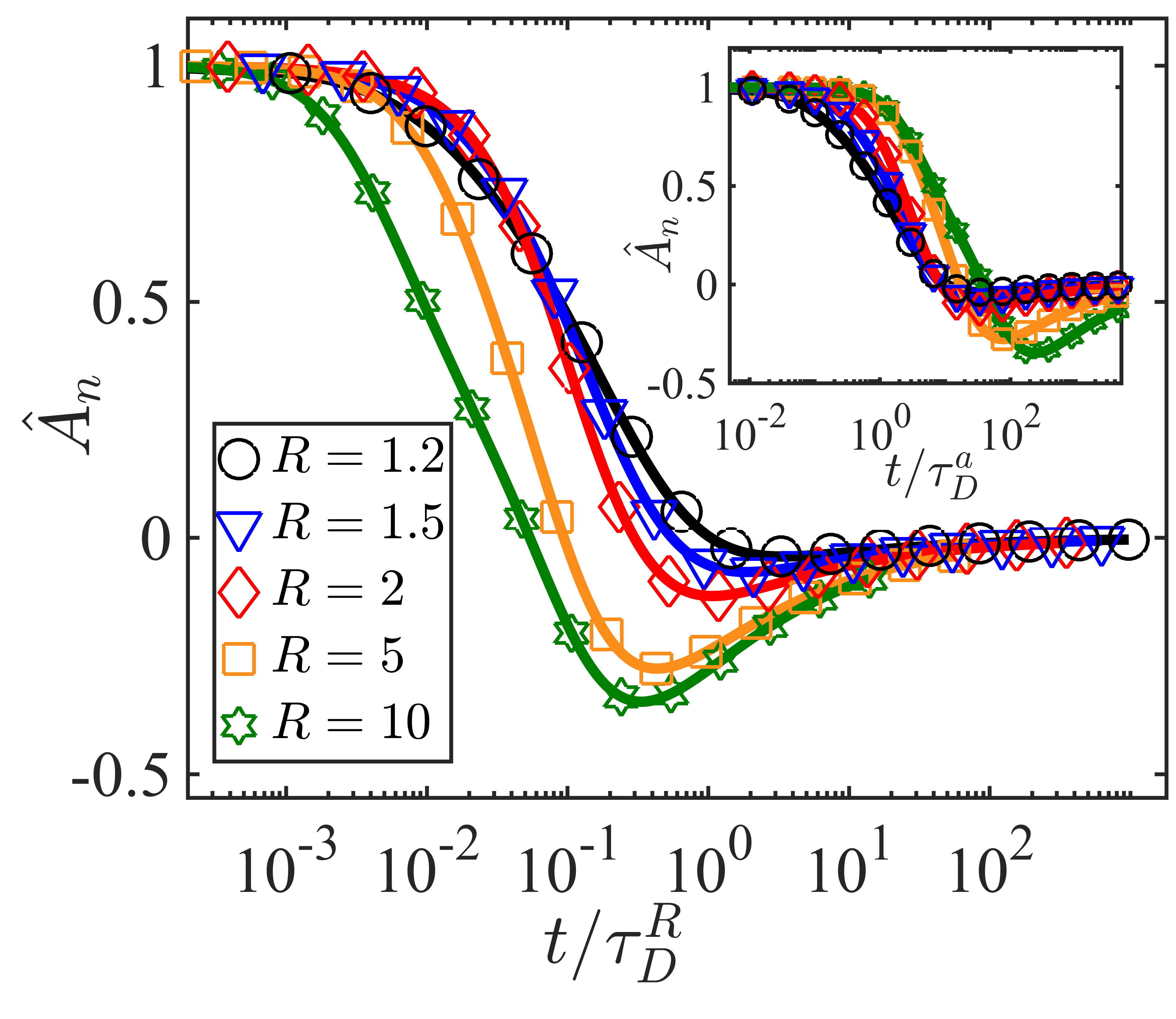}
         \caption*{}
     \end{subfigure}
     \\
     \begin{subfigure}[t]{0.3\textwidth}
         \centering
         \includegraphics[width=\textwidth]{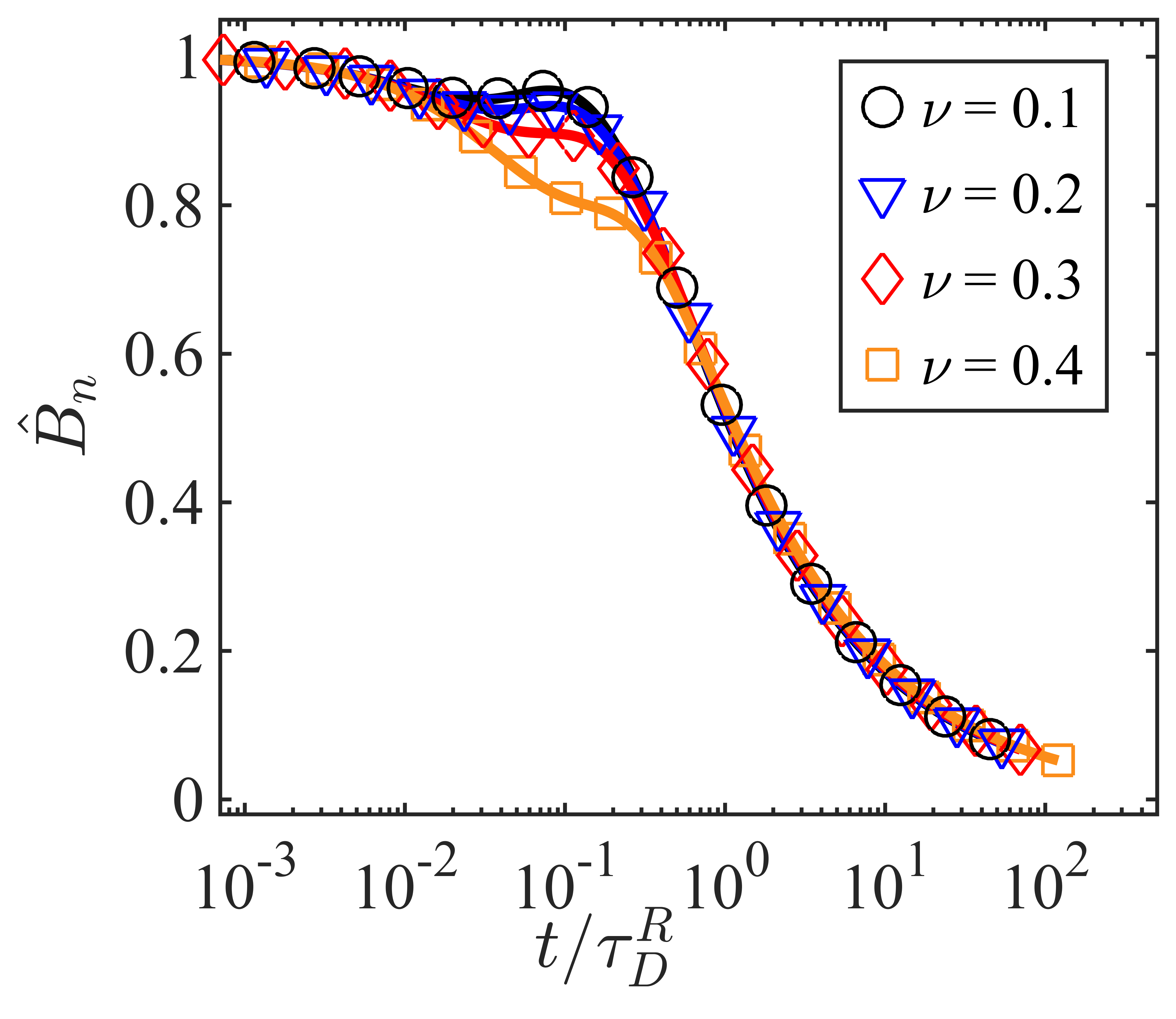}
         \caption{Varying Poisson ratio, $\nu$}
     \end{subfigure}
     \hfill
     \begin{subfigure}[t]{0.3\textwidth}
         \centering
         \includegraphics[width=\textwidth]{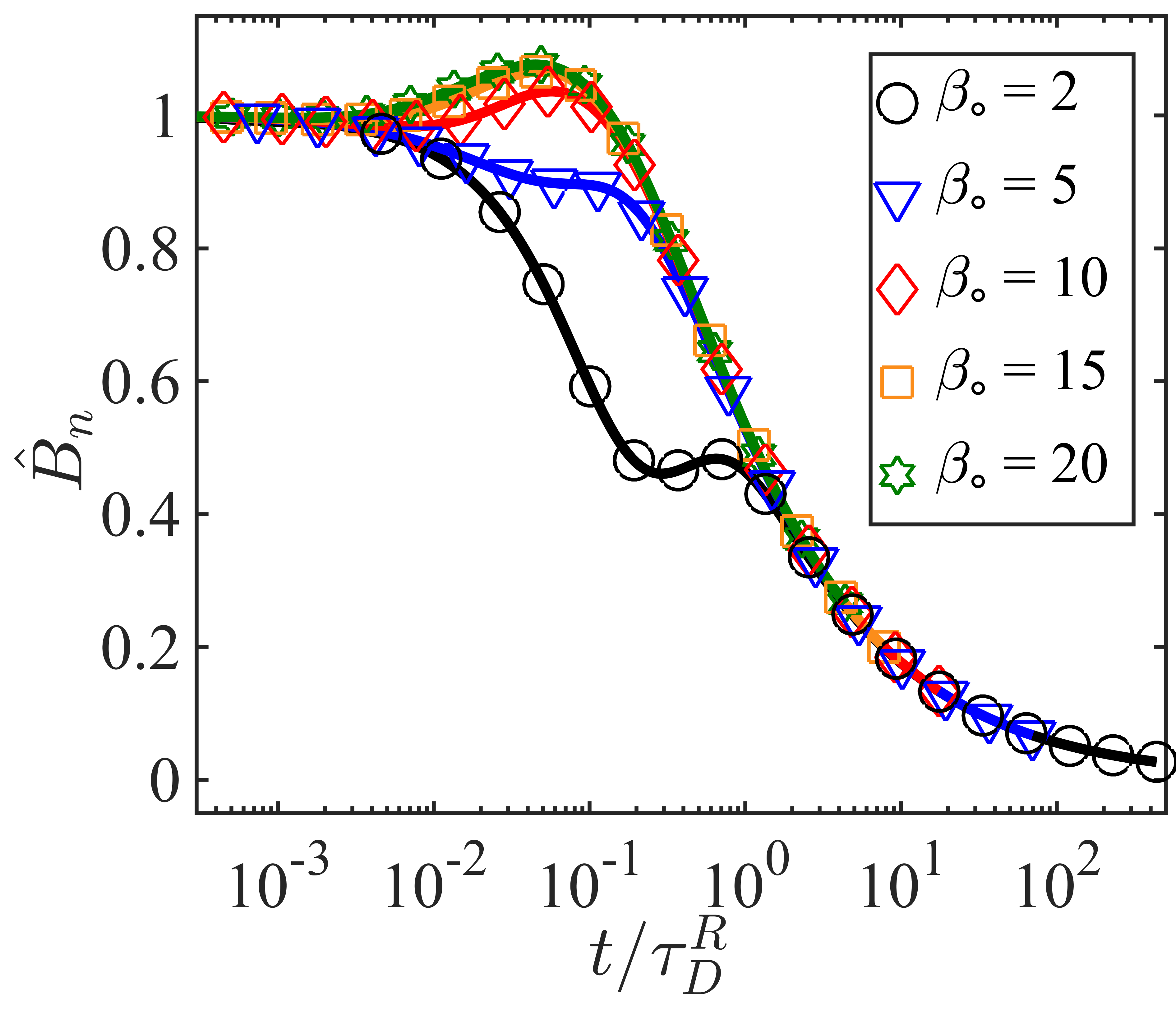}
         \caption{Varying Permeability, $\beta_\circ$}
     \end{subfigure}
     \hfill
     \begin{subfigure}[t]{0.3\textwidth}
         \centering
         \includegraphics[width=\textwidth]{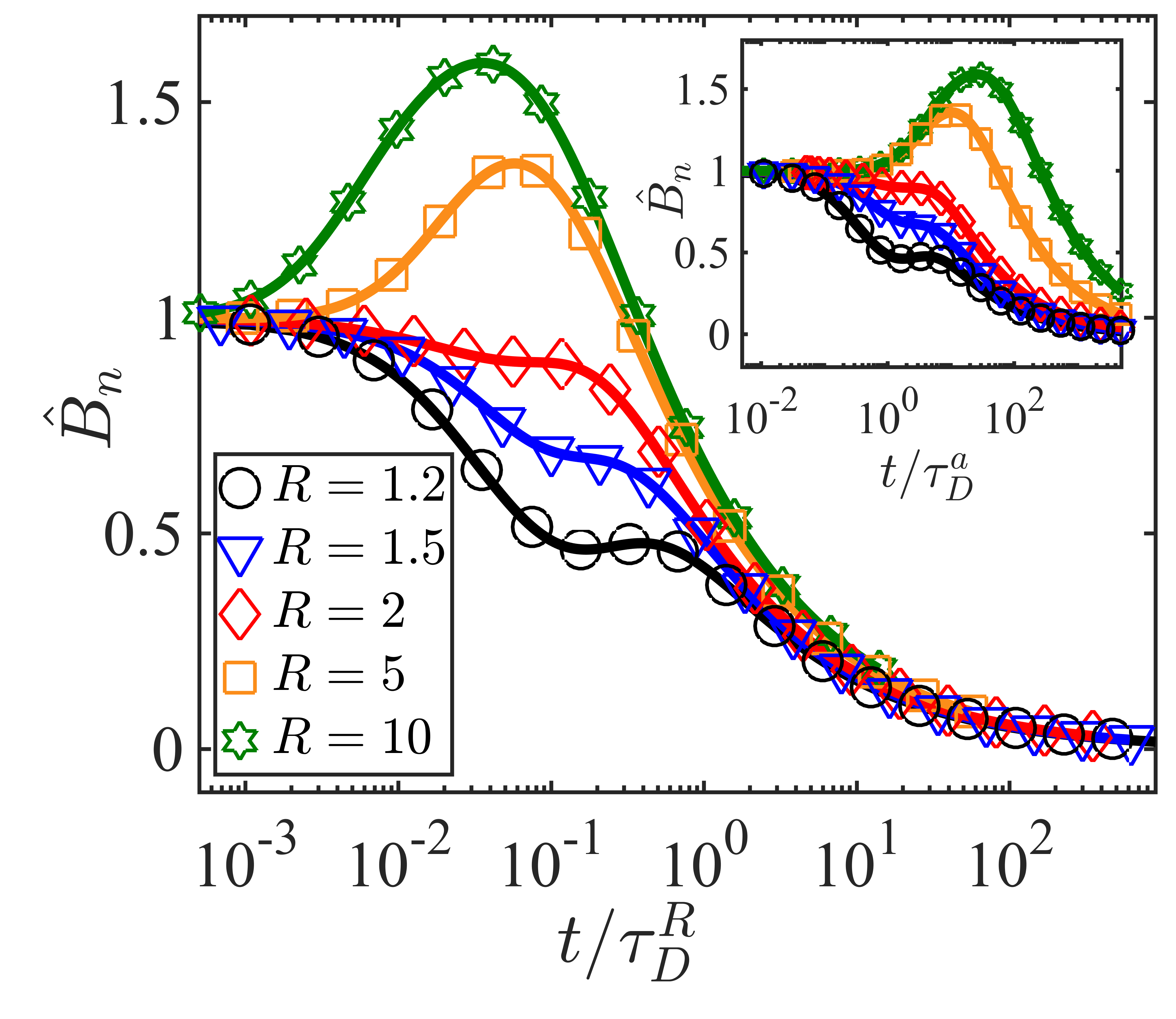}
         \caption{Varying $R$}
     \end{subfigure}
\caption{\emph{Time variations of the network hydrodynamic functions, $\hat{A}_n$ (top row) and $\hat{B}_n$ (bottom row).} (a) Different values of Poisson ratio, $\nu$, while setting $\beta_\circ=5$ and $R=2$. (b) Different values permeability, $\beta_\circ$, while setting $\nu=0.3$ and $R=2$, and (c) different separation distances, $R$, while setting $\nu=0.3$ and $\beta_\circ=5$. 
The time axis is made dimensionless with $\tau_D^R=R^2\alpha_\circ^2\tau_\circ$ in all the main plots. The inset figures in Fig.~\ref{fig4}(c) show the same results as the main figure, but the time axes are made dimensionless with $\tau_D^a=a^2\alpha_\circ^2\tau_\circ$.}
 \label{fig4}
\end{figure}

Figure \ref{fig4} shows the relaxation dynamics of the network functions, $\hat{A}_n$ and $\hat{B}_n$. 
In this case, the solid lines are just showing the detailed predictions and are not showing $\hat{(S_1/\Delta)}$, since $\hat{(S_1/\Delta)}$  values did not match the detailed predictions. 
The time axis is made dimensionless by $\tau_D^{R}$ in all the main plots. Again, when the separation distance is varied, we re-plot the results of Figs. \ref{fig4}(c) in the inset figure, but with the the dimensionless time being defined as $t/\tau_D^a$. 
As it can be seen in Fig. \ref{fig4}(c),  
using $t/\tau_D^R$ gives a better collapse of the relaxation plots at long times and at different $R$ in the main plot, compared to the inset figures, with $t/\tau_D^a$ as x-axis. However, the collapse is not as 
clean as the results for $\hat{A}_f$ and $\hat{B}_f$. 
For both cases of varying permeability and distance (Figs.~\ref{fig4}(b) and \ref{fig4}(c))the plots fail to collapse at shorter times, especially for $\hat{B}_n$, where the shape of the plots qualitatively changes with varying $\beta_\circ$ and $R$. 
All together these results suggest that the relaxation behavior of the network functions $\hat{A}_n$ and $\hat{B}_n$
at long times is distance-dependent and best described by $\tau_D^{R}$. However, the variations at shorter times is a  function of all three timescales ($\tau_\circ$, $\tau_D^a$ and $\tau_D^R$). 

\section{Summary}
\label{sec:summary}
A reciprocal theorem was developed for a two-phase poro-viscoelastic (PVE) material composed of a linear compressible viscoelastic network that is permeated by a linear viscoelastic fluid. The theorem is expressed in Eq.~\eqref{eq:RT2} and, like its counterparts in stokes flow and linear elasticity, is applicable to anisotropic linear materials. 

As an application of the reciprocal theorem, we analytically calculated the net force on a rigid stationary sphere due to point-forces in the fluid and network phases outside of the sphere. We chose the solution to a sphere moving with a prescribed velocity in the PVE medium, which was derived in our earlier work \citep{moradi_shi_nazockdast_2022}, as the auxiliary solution. We showed that the time evolution of the net force due to a point-force in a Newtonian fluid permeating a linear compressible elastic network is independent of the distance between the point-force and the sphere, and primarily determined by the diffusion time of network compressibility over lengths that scale with sphere radius: $\tau_D^a$. In comparison, we found that the relaxation timescale of the net force on the sphere when the point-force is applied to the network is dependent on the distance of the point-force from the sphere, and at long times it is primarily determined by the diffusion time of network compressibility over lengths that scale with this distance: $\tau_D^R$.

The reciprocal theorem can be applied to a variety of other problems. For example, the analysis involving calculating the net force on the sphere due to external point-forces in both phases  can be readily extended to compute (a) the net stress (including pressure) on the sphere due to the same forces, (b) the net force on the sphere due to force dipoles on both phases, and (c) the stress on the sphere due to linear isotropic and shear displacement gradients in both phases. 
When combined with method reflection \citep{happel2012low}, these results can be used to develop the mobility tensors for spherical inclusions in the same manner done in Stokes flow microhydrodynamics \citep{happel2012low, kim2013microhydrodynamics}. These formulas can, then, be used to develop pseudo-analytical methods, such as Stokesian Dynamics \citep{fiore2019fast} and its extensions \citep{swan2011modeling} for computing the dynamics of a large assembly of particles. Furthermore, with some extra work these results can be extended from rigid spheres to viscous or poroelastic inclusions with more complex boundary conditions. These extensions can be useful for studying cell migration in tissues and the extracellular matrix or the dynamics of microswimmers in poroelastic media. 
Another immediate application of our reciprocal theorem is to develop a boundary integral formulation of equations of motion in the same manner as Stokes flow \citep{pozrikidis1992boundary, pozrikidis2002practical}. Boundary integral methods can be used for fast and accurate simulations of the displacement and flow fields in the network and fluid phase containing many inclusions with complex geometry and dynamics. These numerical methods can be used in a wide range of problems in cell mechanics.

\appendix

\section{General solutions for axisymmetric spherical geometry}
\label{App0}
For the special case of the axisymmetric spherical geometry, the general solutions for the fluid and network velocity fields are  \citep{moradi2022general} (${A}_{\ell}^{\pm} $, ${B}_{\ell}^{\pm} $, ${C}_{\ell}^{\pm} $, ${C}_{\ell}^{\prime \,\pm} $, ${D}_{\ell}^{\pm} $ and ${E}_{\ell}^{\pm} $ are constant coefficients):
\begin{subequations}
\begin{align}
& v_{r,f} = \frac{1}{\eta + G} \Bigg[  \sum_{\ell =0 }^{\infty} \Bigg\{ {A}_{\ell}^{\pm} \begin{pmatrix}
 \ell r^{\ell -1} \\ -(\ell +1) r^{-\ell -2} 
 \end{pmatrix} + {B}_{\ell}^{\pm} \frac{\ell (\ell +1)  G}{\beta r}  \begin{pmatrix}
{\mathsf{i}}_{\ell } (\beta r) \\  {\mathsf{k}}_{\ell } (\beta r) %
\end{pmatrix}   \label{eq14a} \\
 & \qquad\qquad + D_{\ell }^{\pm} \begin{pmatrix}
r^{\ell -1} \\ r^{-\ell -2}
\end{pmatrix} \begin{pmatrix} 
\frac{\ell}{2(2\ell +3)} r^2 - \frac{G}{{\eta} {\beta}^2} \ell \\
\frac{\ell +1}{2(2\ell -1)} r^2 + \frac{ G}{ {\eta} {\beta}^2} (\ell +1)   
\end{pmatrix}  \Bigg\} P_{\ell} (\cos\theta)  \Bigg]  ,    \nn  \\
& v_{\theta ,f} =  \frac{1}{\eta + G} \Bigg[  \sum_{\ell =0 }^{\infty} \Bigg\{ {A}_{\ell}^{\pm}  \begin{pmatrix}
  r^{\ell -1} \\ r^{-\ell -2} 
 \end{pmatrix} + {B}_{\ell}^{\pm} \frac{ G}{\beta } \Big[ \frac{\mathrm{d}}{\mathrm{d}r} \begin{pmatrix}
{\mathsf{i}}_{\ell } (\beta r) \\  {\mathsf{k}}_{\ell } (\beta r) 
\end{pmatrix} + \frac{1}{r}  \begin{pmatrix}
{\mathsf{i}}_{\ell } (\beta r) \\  {\mathsf{k}}_{\ell } (\beta r) 
\end{pmatrix}  \Big]   \label{eq14b}  \\
& \qquad\qquad  + D_{\ell }^{\pm} \begin{pmatrix}
r^{\ell -1} \\ r^{-\ell -2}
\end{pmatrix} \begin{pmatrix}
\frac{\ell +3}{2(\ell +1)(2\ell +3)} r^2 - \frac{ G}{{\eta}{\beta}^2}  \\
\frac{2- \ell}{2 \ell (2\ell -1)} r^2 - \frac{ G}{{\eta}{\beta}^2} 
\end{pmatrix}       \Bigg\} \frac{\mathrm{d}}{\mathrm{d}\theta} P_{\ell} (\cos\theta) \Bigg] , \nn \\
&  v_{\varphi ,f} =   \frac{-1}{\eta +G} \Bigg[  \sum_{\ell =0 }^{\infty} \Bigg\{ C_{\ell}^{\pm} \begin{pmatrix}
 r^{\ell} \\ r^{-\ell -1}
\end{pmatrix}  + G \, {C}_{\ell }^{ \prime \,  \pm}  \begin{pmatrix}
{\mathsf{i}}_{\ell } (\beta r) \\  {\mathsf{k}}_{\ell } (\beta r) 
\end{pmatrix}  \Bigg\}  \frac{\mathrm{d}}{\mathrm{d}\theta} P_{\ell} (\cos \theta) \Bigg] ,  \label{eq14c} 
\end{align}
\end{subequations}
\begin{subequations}
\begin{align}
&  v_{r,n} = \frac{1}{\eta + G} \Bigg[  \sum_{\ell =0 }^{\infty} \Bigg\{ {A}_{\ell}^{\pm}  \begin{pmatrix}
 \ell r^{\ell -1} \\ -(\ell +1) r^{-\ell -2} 
 \end{pmatrix} - {B}_{\ell}^{\pm} \frac{\ell (\ell +1) \eta}{\beta r}  \begin{pmatrix}
{\mathsf{i}}_{\ell } (\beta r) \\  {\mathsf{k}}_{\ell } (\beta r) 
\end{pmatrix}     \label{eq15a}  \\
& \qquad \qquad  + D_{\ell }^{\pm} \begin{pmatrix}
r^{\ell -1} \\ r^{-\ell -2}
\end{pmatrix} \begin{pmatrix} 
\frac{\ell}{2(2\ell +3)} r^2 + \frac{1}{{\beta}^2} \ell \\
\frac{\ell +1}{2(2\ell -1)} r^2 - \frac{1}{{\beta}^2} (\ell +1)   
\end{pmatrix}  -  \frac{  \eta + G }{{\alpha}^2} {E}_{\ell}^{\pm} \frac{\mathrm{d}}{\mathrm{d}r}  
 \begin{pmatrix}
{\mathsf{i}}_{\ell } ( {\alpha} r) \\  {\mathsf{k}}_{\ell } ( {\alpha} r) 
\end{pmatrix} \Bigg\} P_{\ell} (\cos\theta) \Bigg] , \nn  \\
&  v_{\theta ,n} =  \frac{1}{\eta + G} \Bigg[  \sum_{\ell =0 }^{\infty} \Bigg\{ {A}_{\ell}^{\pm}   \begin{pmatrix}
  r^{\ell -1} \\ r^{-\ell -2} 
 \end{pmatrix} - {B}_{\ell}^{\pm} \frac{\eta}{\beta } \Big[ \frac{\mathrm{d}}{\mathrm{d}r} \begin{pmatrix}
{\mathsf{i}}_{\ell } (\beta r) \\  {\mathsf{k}}_{\ell } (\beta r) 
\end{pmatrix} + \frac{1}{r}  \begin{pmatrix}
{\mathsf{i}}_{\ell } (\beta r) \\  {\mathsf{k}}_{\ell } (\beta r) 
\end{pmatrix}  \Big]   \label{eq15b}   \\
 & \qquad \qquad  + D_{\ell }^{\pm} \begin{pmatrix}
r^{\ell -1} \\ r^{-\ell -2}
\end{pmatrix} \begin{pmatrix}
\frac{\ell +3}{2(\ell +1)(2\ell +3)} r^2 + \frac{1}{{\beta}^2}  \\
\frac{2- \ell}{2 \ell (2\ell -1)} r^2 + \frac{1}{{\beta}^2} 
\end{pmatrix}       -  \frac{ \eta + G }{{\alpha}^2} {E}_{\ell}^{\pm} \frac{1}{r} 
 \begin{pmatrix}
{\mathsf{i}}_{\ell } ( {\alpha} r) \\  {\mathsf{k}}_{\ell } ( {\alpha} r) 
\end{pmatrix} \Bigg\} \frac{\mathrm{d}}{\mathrm{d}\theta} P_{\ell} (\cos\theta) \Bigg] , \nn \\
&  v_{\varphi ,n} =   \frac{-1}{\eta +G} \Bigg[  \sum_{\ell =0 }^{\infty} \Bigg\{ C_{\ell}^{\pm} \begin{pmatrix}
 r^{\ell} \\ r^{-\ell -1}
\end{pmatrix}  - \eta \, {C}_{\ell }^{ \prime \, \pm}  \begin{pmatrix}
{\mathsf{i}}_{\ell } (\beta r) \\  {\mathsf{k}}_{\ell } (\beta r) 
\end{pmatrix}  \Bigg\}  \frac{\mathrm{d}}{\mathrm{d}\theta} P_{\ell} (\cos \theta) \Bigg]    \label{eq15c} . 
\end{align}
\end{subequations}
Here, $P_{\ell} (x)$ are the Legendre polynomials, and  ${\mathsf{i}}_{\ell} (x )  $, and ${\mathsf{k}}_{\ell} (x)  $ are the modified spherical Bessel functions of first and second kind, respectively \citep{arfken1999mathematical}. 
For brevity, we have presented these solutions in the array format $()^\pm$. The first row contains the internal solutions (the functions are finite when
$r\to 0$ and are unbounded as $r\to \infty$), and the second row contains the external solutions
(the functions decay to zero as $r\to \infty$ and are singular as $r\to 0$). 
Also, the pressure is:
\begin{eqnarray}
&& p = \sum_{\ell=0}^\infty  \Bigg\{ D_{\ell }^{\pm} \begin{pmatrix}
r^{\ell} \\ r^{-\ell -1}
\end{pmatrix} - \big(  \lambda + 2 G \big) E_{\ell }^{\pm}   \begin{pmatrix}
{\mathsf{i}}_{\ell } ( \alpha r) \\  {\mathsf{k}}_{\ell } ( \alpha r) 
\end{pmatrix} \Bigg\} P_{\ell } (\cos\theta ). \label{Pressure}
\end{eqnarray}

\bibliographystyle{unsrt}
\bibliography{MybibE}

\end{document}